\documentclass[12pt]{article}

\usepackage{amsfonts}
\usepackage[centertags]{amsmath}
\usepackage{amssymb}
\usepackage[verbose,colorlinks=true,naturalnames=true,linkcolor=blue,]{hyperref}
\usepackage{ulem}
\usepackage{xcolor}

\def\ba{\begin{array}}
\def\ea{\end{array}}
\def\be{\begin{equation}}
\def\ee{\end{equation}}
\def\bea{\begin{eqnarray}}
\def\eea{\end{eqnarray}}
\usepackage[verbose,colorlinks=true,naturalnames=true,linkcolor=blue,]{hyperref}

\newcounter{rown}

\begin{document}

\title{All basic quantizations of $D=3$, $N=1$ \\ Lorentz supersymmetry} 

\author{  V.N. Tolstoy \\
Lomonosov Moscow State University,\\
Skobeltsyn Institute of Nuclear Physics, \\
Moscow 119991, Russian Federation}
\date{}
\maketitle

\begin{abstract}
By the supersymmetrization of a simple algebraic technique proposed in \cite{LuTo2017} we obtain the complete classification of all basic (nonisomorphic) quantum deformations for the orthosymplectic Lie superalgebra $\mathfrak{osp}(1|2;\mathbb{C})$ and its pseudoreal and real forms in terms of the classical $r$-matrices. In particular, we prove that pseudoreal compact form has only one quantum deformation (standart $q$-analog), and the $D=3$, $N=1$ Lorentz supersymmetry, which is the non-compact real form of $\mathfrak{osp}(1|2;\mathbb{C})$, has four different Hopf-algebraic quantum deformations: two standard $q$-analogs, and two (Jordanian and super-Jordanian) twist deformations. All basic Hopf-algebraic quantum deformations are presented in the explicit form.
\end{abstract}

\setcounter{equation}{0} 
\section{Introduction}

The search for quantum gravity is linked with studies of noncommutative space-times and quantum deformations of space-time symmetries. The considerations of simple dynamical models in quantized gravitational background (see e.g. \cite{FrLi2006,CiKoPrRo2016}) indicate that the presence of quantum gravity effects generates noncommutativity of space-time coordinates, and as well the  well-known  Lie-algebraic space-time symmetries (e.g. Euclidean, Lorentz, Poincare) are modified into respective quantum symmetries, described by noncocommutative Hopf algebras, named quantum deformations or quantum groups \cite{Dr1986}. Therefore, studding all aspects of the quantum deformations in details is an important issue in particular in the search of quantum gravity models and their quantum superextensions (SUGRA).

For classifications, constructions and applications of quantum Hopf deformations of an universal enveloping algebra $U(\mathfrak{g})$ of a Lie superalgebra $\mathfrak{g}$ ($\mathfrak{g}=\mathfrak{g}_{\bar{0}}\oplus\mathfrak{g}_{\bar{1}}$, as a linear space, with a linear $\mathbb{Z}_2$-grading function $\deg(\cdot)$: $\deg(\mathfrak{g}_{a})=a\in\{\bar{0},\bar{1}\}$) Lie bisuperalgebras ($\mathfrak{g},\delta$) play an essential role (in analogy to the case of Lie algebras (see e.g. \cite{Dr1986, EtKa1996} and \cite{ChPr1994, Ma1995})), where the {\it cobracket} $\delta$ is a linear super-skew-symmetric map:
\begin{eqnarray}\label{in1}
\mathfrak{g}\rightarrow\mathfrak{g}\wedge\mathfrak{g}=\mathfrak{g}_{\bar{0}}\wedge\mathfrak{g}_{\bar{0}}\oplus\mathfrak{g}_{\bar{1}}\wedge\mathfrak{g}_{\bar{1}}\oplus\mathfrak{g}_{\bar{0}}\wedge\mathfrak{g}_{\bar{1}},
\end{eqnarray}
which conserves the grading function $\deg(\cdot)$: 
\begin{eqnarray}\label{in2}
\delta(\mathfrak{g}_{\bar{0}})\in\mathfrak{g}_{\bar{0}}\wedge\mathfrak{g}_{\bar{0}}\oplus\mathfrak{g}_{\bar{1}}\wedge\mathfrak{g}_{\bar{1}},\quad\delta(\mathfrak{g}_{\bar{1}})\in\mathfrak{g}_{\bar{0}}\wedge\mathfrak{g}_{\bar{1}}.
\end{eqnarray}
provided that 
\begin{eqnarray}\label{in3}
\deg(\mathfrak{g}_{a}\wedge\mathfrak{g}_{b})=\deg(\mathfrak{g}_{a})+\deg(\mathfrak{g}_{b})=a+b=c\mod2 
\end{eqnarray}\\[-5pt]
for $a,b,c\in\{\bar{0},\bar{1}\}$. In the general case $x=x_{\bar{0}}+x_{\bar{1}}$ we use the linearity of $\delta(x)$ in the argument $x$. It should be noted also that the superskew-symmetric bilinear space $\mathfrak{g}\wedge\mathfrak{g}$ is defined as follows:
\begin{eqnarray}\label{in4}
x_a\wedge y_b:= x_a\otimes y_b - (-1)^{ab} y_b\otimes x_a
\end{eqnarray}
for all homogeneous elements $x_a\in g_a$, $y_b\in g_b$ ($a,b\in\{\bar{0},\bar{1}\}$). In case of general elements $x=x_{\bar{0}}+x_{\bar{1}}$ and $y=y_{\bar{0}}+y_{\bar{1}}$ we use the bilinearity of $x\wedge y$.

Moreover the cobracket $\delta$ satisfies the relations consisted with the superbracket $[\![\cdot,\cdot]\!]$ in the Lie superalgebra $\mathfrak{g}$:
\begin{eqnarray}
&&\begin{array}{rcl}\label{in5}
\delta([\![x,y]\!])\!\!&=\!\!&[\![\Delta_{0}(x),\delta(y)]\!]+[\![\delta(x),\Delta_{0}(y)]\!]
\\[5pt]
&=\!\!&[x,\delta(y)]_{S}+[\delta(x),y]_{S}\quad (x,y\in\mathfrak{g}),
\end{array}
\\[5pt]
&&\begin{array}{rcl}\label{in6} 
(\delta\otimes\mathop{\rm id})\delta(x_a)+{\rm grcycle}\,=\,0,
\end{array}
\end{eqnarray}
for any homogeneous element $x_a\in g_a,\;a\in\{\bar{0},\bar{1}\}$. Here $\Delta_{0}(\cdot)$ is a trivial (non-deformed) coproduct 
\begin{eqnarray}\label{in7}
\Delta_{0}(x)=x\otimes1+1\otimes x, 
\end{eqnarray}
and the symbol $[\cdot,\cdot]_{S}$ means the graded Schouten bracket (\ref{rm2}). 
The first relation (\ref{in5}) is a condition of the 1-cocycle and the second one (\ref{in6}) is the co-Jacobi identity. The Lie bisuperalgebra ($\mathfrak{g},\delta$) is a correct infinitesimalization of the quantum Hopf deformation  of $U(\mathfrak{g})$ and the operation $\delta$ is an infinitesimal part of difference between a coproduct $\Delta$ and an opposite coproduct $\tilde{\Delta}$ in the Hopf algebra, $\delta(x)=h^{-1}(\Delta-\tilde{\Delta})\mod h$, where $h$ is a deformation parameter. Any two Lie bialgebras ($\mathfrak{g},\delta$) and ($\mathfrak{g},\delta'$) are isomorphic (equivalent) if they are connected by a $\mathfrak{g}$-automorphism $\varphi$ satisfying the condition 
\begin{eqnarray}\label{in8}
\delta(x)\,=\,(\varphi\otimes\varphi)\delta'(\varphi^{-1}(x)) 
\end{eqnarray}
for any $x\in\mathfrak{g}$.

Of our special interest here are the quasitriangular Lie bisuperalgebras ($\mathfrak{g},\delta_{(r)}$):=($\mathfrak{g},\delta,r$), where the cobracket $\delta_{(r)}$ is given by the classical $r$-matrix $r\in\mathfrak{g}\wedge\mathfrak{g}$ as follows: 
\begin{eqnarray}\label{in9}
\delta_{(r)}(x)\,=\,[\![\Delta_{0}(x),r]\!]\,=\,[x,r]_{S}.
\end{eqnarray}
Because the co-bracket $\delta_{(r)}$ conserves the grading value then from (\ref{in9}) we see that the $r$-matrix $r$ is even one, $\deg(r)=\bar{0}$, i.e.
\begin{eqnarray}\label{in10}
r\;\in\;\mathfrak{g}_{\bar{0}}\wedge\mathfrak{g}_{\bar{0}}\oplus\mathfrak{g}_{\bar{1}}\wedge\mathfrak{g}_{\bar{1}}. 
\end{eqnarray}
Moreover it is easy to see also from (\ref{in8}) and (\ref{in9}) that \textit{two quasitriangular Lie bisuperalgebras ($\mathfrak{g},\delta_{(r)}$) and ($\mathfrak{g}, \delta_{(r')}$) are isomorphic iff the classical $r$-matrices $r$ and $r'$ are isomorphic, i.e.} 
\begin{eqnarray}\label{in11}
(\varphi\otimes\varphi)r'=r. 
\end{eqnarray}
Therefore for a classification of all nonequivalent quasitriangular Lie bisuperalgebras ($\mathfrak{g},\delta_{(r)}$) of the given Lie superalgebra $\mathfrak{g}$ we need to find all nonequivalent (nonisomorphic) classical $r$-matrices. Because nonequivalent quasitriangular Lie bisuperalgebras uniquely determine non-equivalent quasitriangular quantum deformations (Hopf algebras) of $U(\mathfrak{g})$ therefore the classification of all nonequivalent quasitriangular Hopf superalgebras is reduced to the classification of all nonequivalent classical $r$-matrices. 

Let $\mathfrak{g}^{*}:=(\mathfrak{g},*)$ be a real or pseudoreal\footnote{See Sect. 2, where a pseudoreality condition for the superalgebra $\mathfrak{osp}^{*}(1|\mathfrak{su}(2))$ is considered.} form of a classical complex Lie superalgebra $\mathfrak{g}$, where $*$ is an antilinear involutive or semiinvolutive  antiautomorphism of $\mathfrak{g}$, then \textit{the bisuperalgebra $(\mathfrak{g}^{*},\delta_{(r)})$ is real iff the classical $r$-matrix $r$ is $*$-anti-real ($*$-anti-Hermitian)}.\footnote{All bialgebras over the simple complex and real Lie algebras are quasitriangular, due to Whitehead lemma (see e.g. \cite{Ja1979}). It is more likely that the Whitehead lemma is valid also for all classical Lie superalgebras.} Indeed, the condition of $*$-reality for the bisuperalgebra $(\mathfrak{g}^{*},\delta)$ means that 
\begin{eqnarray}\label{in12}
\delta(x)^{*\otimes*}=\delta(x^{*}).
\end{eqnarray}
Applying this condition to the relations (\ref{in9}) we obtain that 
\begin{eqnarray}\label{in13}
r^{*\otimes*}=-r,
\end{eqnarray}
i.e. the $r$-matrix $r$ is $*$-anti-Hermitian.

Recently in the paper \cite{LuTo2017} there were investigated the quantum deformations of the complex Lie algebra $\mathfrak{sl}(2;\mathbb{C})\simeq\mathfrak{o}(3;\mathbb{C})$ and its real forms $\mathfrak{su}(2)\simeq\mathfrak{o}(3)$, $\mathfrak{sl}(2;\mathbb{R})\simeq\mathfrak{o}(2,1)$ and $\mathfrak{su}(1,1)\simeq\mathfrak{o}(2,1)$. Namely, firstly it was obtained the complete classifications of the nonequivalent (nonisomorphic) classical $r$-matrices (bialgebras) for all these Lie algebras and then Hopf deformations corresponding to these bialgebras were presented in explicit form. In particular, it was shown that $D=3$ Lorentz symmetry $\mathfrak{o}(2,1)\,(\simeq\mathfrak{sl}(2;\mathbb{R})\simeq\mathfrak{su}(1,1)$) has two standard $q$-deformations and one Jordanian.

In this work we would like to present some superanalog of these results, namely we first give the complete classifications of the nonequivalent (nonisomorphic) classical $r$-matrices for complex Lie superalgebra $\mathfrak{osp}(1|2;\mathbb{C})$ (which is a minimal supersymmetric extension of the Lie algebra $\mathfrak{sl}(2;\mathbb{C})$), and its pseudoreal $\mathfrak{osp}^{*}(1|\mathfrak{su}(2))$\footnote{It should be noted that the compact forms for all superalgebras of the classical series $\mathfrak{osp}(1|2m;\mathbb{C})$ ($m=1,2,\ldots$) are  pseudoreal.}  and real $\mathfrak{osp}^{\dag}(1|\mathfrak{sl}(2;\mathbb{R}))$, $\mathfrak{osp}^{\dag}(1|\mathfrak{su}(1,1))$ forms. In particular, it will be shown that $N=1$, $D=3$ Lorentz supersymmetry, 
\begin{eqnarray}\label{in14}
\mathfrak{osp}^{\dag}(1|\mathfrak{o}(2,1))\simeq\mathfrak{osp}^{\dag}(1|\mathfrak{sl}(2;\mathbb{R}))\simeq\mathfrak{osp}^{\dag}(1|\mathfrak{su}(1,1)), 
\end{eqnarray}
has two nonequivalent standard $q$-deformations and two nonequivalent (Jordanian and super-Jordanian) twist deformations. Moreover it will be shown that each real form $\mathfrak{osp}^{*}(1|\mathfrak{o}(3))$ and $\mathfrak{osp}^{\dag}(1|\mathfrak{o}(2,1))$ has four variants of the deformation in accordance with how the super-involutions ${*}$ and $\dag$ are extended on all universal enveloping superalgebra $U(\mathfrak{osp}(1|2;\mathbb{C}))$ and its tensor square (see Section 4).
The isomorphic Lie superalgebras (\ref{in14}) and their quantum deformations play very important role in physics as well as in mathematical considerations, so the structure of these deformations should be understood with full clarity. It should be noted also that the importance of $\mathfrak{osp}^{\dag}(1|\mathfrak{o}(2,1))$ and its deformations follows also from the unique role of the $\mathfrak{osp}^{\dag}(1|\mathfrak{o}(2,1))$ superalgebra as the lowest-dimensional rank one noncompact simple classical Lie superalgebra.

In this paper we investigate the quantum deformations of $D=3$, $N=1$ Lorentz supersymmetry. Firstly, following \cite{LuTo2017,BoLuTo2016}, we obtain the complete classifications of the nonequivalent (nonisomorphic) classical $r$-matrices for complex Lie superalgebra $\mathfrak{osp}(1|2;\mathbb{C})$ and its pseudoreal and real forms $\mathfrak{osp}^{*}(1|\mathfrak{su}(2))$, $\mathfrak{osp}^{\dag}(1|\mathfrak{sl}(2;\mathbb{R}))$ and $\mathfrak{osp}^{\dag}(1|\mathfrak{su}(1,1))$ with the help of explicit formulas for the automorphisms of these Lie superalgebras in terms of the Cartan-Weyl bases. In the case of $\mathfrak{osp}(1|2;\mathbb{C})$ there are three nonequivalent classical $r$-matrices - one standard form and two Jordanian and super-Jordanian ones. For the pseudoreal superalgebra $\mathfrak{osp}^{*}(1|\mathfrak{su}(2))$  there is only the standard classical $r$-matrix. For the real case $\mathfrak{osp}^{\dag}(1|\mathfrak{su}(1,1))$  we obtained four noneqvivalent $r$-matrices - standard, quasi-standard, quasi-Jordanian and quasi-super-Jordanian ones. In the real case of $\mathfrak{osp}^{\dag}(1|\mathfrak{sl}(2;\mathbb{R}))$ we find also four nonequivalent $r$-matrices - standard, quasi-standard, Jordanian and super-Jordanian ones. Then using isomorphisms $\mathfrak{osp}^{\dag}(1|\mathfrak{o}(2,1))\simeq\mathfrak{osp}^{\dag}(1|\mathfrak{su}(1,1))\simeq\mathfrak{osp}^{\dag}(1|\mathfrak{sl}(2;\mathbb{R}))$ we express these $r$-matrices in terms of the Cartesian basis of the $D=3$, $N=1$ Lorentz superalgebra $\mathfrak{osp}^{\dag}(1|\mathfrak{o}(2,1))$ and we see that two systems with four $r$-matrices for $\mathfrak{osp}^{\dag}(1|\mathfrak{su}(1,1))$ and $\mathfrak{osp}^{\dag}(1|\mathfrak{sl}(2;\mathbb{R}))$ superalgebras coincides. Thus we obtain that the isomorphic Lie algebras $\mathfrak{osp}^{\dag}(1|\mathfrak{su}(1,1))$ and $\mathfrak{osp}^{\dag}(1|\mathfrak{sl}(2;\mathbb{R}))$ have the isomorphic systems of their quasitriangular Lie bealgebras. In the case of $\mathfrak{osp}^{\dag}(1|\mathfrak{o}(2,1))$ we obtain that the $D=3$, $N=1$ Lorentz superalgebra has two standard $q$-deformations and two Jordanian and super-Jordanian ones. These Hopf deformations are presented in explicit form in terms of the quantum Cartan-Weyl generators for the quantized universal enveloping algebras of $\mathfrak{osp}^{\dag}(1|\mathfrak{su}(1,1))$ and $\mathfrak{osp}^{\dag}(1|\mathfrak{sl}(2;\mathbb{R}))$. 
It should be noted that the full list of the nonequivalent classical $r$-matrices for the complex Lie superalgebra $\mathfrak{osp}(2|1;\mathbb{C})$ was obtained previously in \cite{JuSo1998} where the authors used a technique of computer algebraic computations. However, the complete list of the nonequivalent classical $r$-matrices and their Hopf quantizations for the real $D=3$, $N=1$ Lorentz superalgebra $\mathfrak{osp}^{\dag}(1|\mathfrak{o}(2,1))$ has not been presented in the literature, but some examples of such $r$-matrices and their quantizations were already considered (for example, see \cite{Ku1988,KuRe1989,LuNo1992,BoLuTo2003}).

The isomorphic Lie superalgebras $\mathfrak{osp}^{\dag}(1|\mathfrak{o}(2,1))$, $\mathfrak{osp}^{\dag}(1|\mathfrak{sl}(2;\mathbb{R}))$, $\mathfrak{osp}^{\dag}(1|\mathfrak{su}(1,1))$ and their quantum deformations play very important role in physics as well as in mathematical considerations, so the structure of these deformations should be understood with full clarity. For example, since the real superalgebra $\mathfrak{osp}^{\dag}(1|\mathfrak{sl}(2;\mathbb{R}))$ can be used as $D=1$, $N=1$ superconformal symmetry then the obtained results in this case allow us to interpret as deformations of $N=1$ superconformal mechanics \cite{BoLuTo2003}. 
In mathematics and mathematical physics the importance of $\mathfrak{osp}^{\dag}(1|\mathfrak{o}(2,1))$ and its deformations follows also from the unique role of the $\mathfrak{osp}^{\dag}(1|\mathfrak{o}(2,1))$ algebra as the lowest-dimensional rank one noncompact simple Lie superalgebra, endowed only with unitary infinite-dimensional representations. 

The plan of this paper is the following. In Sect.~2 we consider the complex Lie superalgebra $\mathfrak{osp}(1|2;\mathbb{C})$ and its all real forms: $\mathfrak{osp}^{*}(1|\mathfrak{o}(3))\simeq\mathfrak{osp}^{*}(1|\mathfrak{su}(2))$, $\mathfrak{osp}^{\dag}(1|\mathfrak{o}(2,1))\simeq\mathfrak{osp}^{\dag}(1|\mathfrak{su}(1,1))\simeq\mathfrak{osp}^{\dag}(1|\mathfrak{sl}(2;\mathbb{R}))$. In Sect.~3 we classify all classical $r$-matrices for the complex Lie superalgebra $\mathfrak{osp}(1|2;\mathbb{C})$ and in Sect.~4  all classical $r$-matrices for its real forms: $\mathfrak{osp}^{*}(1|\mathfrak{su}(2))$, $\mathfrak{osp}^{\dag}(1|\mathfrak{su}(1,1))$ and $\mathfrak{osp}^{\dag}(1|\mathfrak{sl}(2;\mathbb{R}))$. In Sect.~5 we provide the explicit isomorphisms between the $\mathfrak{osp}^{\dag}(1|\mathfrak{su}(1,1))$, $\mathfrak{osp}^{\dag}(1|\mathfrak{sl}(2;\mathbb{R}))$ and $\mathfrak{osp}^{\dag}(1|\mathfrak{o}(2,1))$ bialgebras. In Sect.~6 all four Hopf-algebraic quantizations (explicit quantum deformations) of the real $D=3$ Lorentz supersymmetry are presented in detail: quantized bases, coproducts and universal $R$-matrices are given. In Sect.~7 we present short summary and outlook. 
\setcounter{equation}{0}
\section{Complex Lie superalgebra $\mathfrak{osp}(1|2;\mathbb{C})$ and its real forms}
The Lie superalgebra $\mathfrak{osp}(1|2;\mathbb{C})$ is a initial element of infinite orthosymplectic series $\mathfrak{osp}(m|2n;\mathbb{C})$ ($m,n=1,2,\ldots$). Each orthosymplectic  Lie superalgebra $\mathfrak{osp}(m|2n;\mathbb{C})$, as a linear space, is a direct sum of two graded (even and odd) components: $\mathfrak{osp}(m|2n;\mathbb{C})=\mathfrak{osp}(m|2n;\mathbb{C})_{0}\oplus\mathfrak{osp}(m|2n;\mathbb{C})_{1}$, where the even part $\mathfrak{osp}(m|2n;\mathbb{C})_{0}$ is a direct sum of the orthogonal and symplectic Lie algebras: $\mathfrak{osp}(m|2n;\mathbb{C})_{0}=\mathfrak{o}(m;\mathbb{C})\oplus\mathfrak{sp}(2n;\mathbb{C})$, and moreover $[\mathfrak{osp}(m|2n;\mathbb{C})_{0},\mathfrak{osp}(m|2n;\mathbb{C})_{1}]=\mathfrak{osp}(m|2n;\mathbb{C})_{1}$, $\{\mathfrak{osp}(m|2n;\mathbb{C})_{1}$, $\mathfrak{osp}(m|2n;\mathbb{C})_{1}\}=\mathfrak{osp}(m|2n;\mathbb{C})_{0}$. 

In the case of $\mathfrak{osp}(1|2;\mathbb{C})$ the even part has the form $\mathfrak{osp}(1|2;\mathbb{C})_{0}=\mathfrak{o}(1;\mathbb{C})\oplus\mathfrak{sp}(2;\mathbb{C})$ where $\mathfrak{o}(1;\mathbb{C})$ is null-algebra $\mathfrak{o}(1;\mathbb{C})\simeq\{0\}$ and $\mathfrak{sp}(2;\mathbb{C})\simeq\mathfrak{sl}(2;\mathbb{C})\simeq\mathfrak{o}(3;\mathbb{C})$. Let $\{E_{\pm},H\}$ be a Cartan-Weyl (CW) basis of $\mathfrak{sl}(2;\mathbb{C})\simeq\mathfrak{o}(3;\mathbb{C})$ with the standard relations:
\begin{eqnarray}\label{pr1}
&&[H,E_{\pm}]=\pm E_{\pm},\;\;[E_{+},E_{-}]=2H,
\end{eqnarray}
then in the odd two-dimensional space $\mathfrak{osp}(1|2;\mathbb{C})_{1}$ one can choose the basis $(v_{+},v_{-})$ satisfying the relations(see, for example,  \cite{ScNaRi1978,BeTo1981,IvLeZu2004}):
\begin{eqnarray}\label{pr2}
&&[H,v_{\pm}]=\displaystyle\pm\frac{1}{2}v_{\pm},\;\;[E_{\mp},v_{\pm}]=v_{\mp},\;\; [E_{\pm},v_{\pm}]=0,
\\[3pt] \label{pr3}
&&\{v_{\pm},v_{\pm}\}=\displaystyle\pm\frac{1}{2}E_{\pm},\;\; \{v_{+},v_{-}\}=-\frac{1}{2}H.
\end{eqnarray}
The the CW generators $H,E_{\pm}$ of $\mathfrak{sl}(2;\mathbb{C})\simeq\mathfrak{o}(3;\mathbb{C})$ is related with the {\it Cartesian} basis $I_{i}$ $(i=1,2,3)$  as follows: 
\begin{eqnarray}\label{pr4}
&&H=\imath I_{3},\;\; E_{\pm}=\imath I_{1}\mp I_{2}.
\end{eqnarray}
For convenience we set also 
\begin{eqnarray}\label{pr5}
v_{1}:=v_{+},\;\; v_{2}:=v_{-}.
\end{eqnarray}
In terms of the generators $\{I_{i},v_{\alpha}|i=1,2,3;\alpha=1,2\}$ the defining relations (\ref{pr1})--(\ref{pr3}) take the form:
\begin{eqnarray}\label{pr6}
\begin{array}{rcl}
&&[I_{i},I_{j}]=\varepsilon_{ijk}I_{k},\;\; [I_{i},v_{\alpha}]=\displaystyle-\frac{\imath}{2}(\sigma_{i})_{\beta\alpha}v_{\beta},
\\[9pt]
&&\{v_{1},v_{1}\}=\displaystyle\frac{1}{2}(\imath I_{1}-I_{2}),\;\;\{v_{1},v_{2}\}=-\displaystyle\frac{\imath}{2}I_{3},
\\[9pt]
&&\{v_{2},v_{2}\}=\displaystyle-\frac{1}{2}(\imath I_{1}+I_{2}),
\end{array}
\end{eqnarray}
where $\sigma_{i}$, $(i=1,2,3)$ are the $2\times2$ Pauli matrices, and $(\alpha,\beta=1,2)$.

It is well known that the Lie algebra $\mathfrak{o}(3;\mathbb{C})\simeq\mathfrak{sl}(2;\mathbb{C})$, which is a subalgebra of the superalgebra $\mathfrak{osp}(1|2;\mathbb{C})$, has two real forms: compact $\mathfrak{o}(3)\simeq\mathfrak{su}(2)$, and noncompact $\mathfrak{o}(2,1)\simeq\mathfrak{sl}(2;\mathbb{R})\simeq\mathfrak{su}(1,1)$. These real forms of the subalgebra $\mathfrak{o}(3;\mathbb{C})\simeq\mathfrak{sl}(2;\mathbb{C})$ are raised up to the odd part $\mathfrak{osp}_{1}(1|2;\mathbb{C})$, that is to the whole superalgebra $\mathfrak{osp}(1|2;\mathbb{C})$.

I. \textit{The compact pseudoreal superalgebra $\mathfrak{osp}^{*}(1|\mathfrak{o}(3))\simeq\mathfrak{osp}^{*}(1|\mathfrak{su}(2))$}.\\ 
In terms of the generators $\{I_{i},v_{\alpha}|i=1,2,3;\alpha=1,2\}$ this form is defined by the following conjugation:
\begin{eqnarray}\label{pr7}
&&I^{*}_{i}=-I_{i},\;\; v^{*}_{1}=\varepsilon v_{2},\;\; v^{*}_{2}=-\varepsilon v_{1},
\end{eqnarray}
where $\varepsilon=1$ if the conjugation ($^*$) of the Lie superbracket is graded, i.e.
\begin{eqnarray}\label{pr8}
&&[\![x_{a},x_{b}]\!]^*=(-1)^{ab}[\![x_{b}^*,x_{a}^*]\!]\;\; {\rm (graded)},
\end{eqnarray}
and $\varepsilon=\imath$ if the conjugation ($^*$) of the Lie superbracket is not graded, i.e.
\begin{eqnarray}\label{pr9}
&&[\![x_{a},x_{b}]\!]^*=[\![x_{b}^*,x_{a}^*]\!]\;\; {\rm (ungraded)},
\end{eqnarray}
for all homogeneous elements $x_{a}\in\mathfrak{g}_{a}$, $x_{b}\in\mathfrak{g}_{b}$ ($\mathfrak{g}:=\mathfrak{osp}(1|2;\mathbb{C})$). We see that the conjugation ($^*$) prolonged to the odd generators $v_{\alpha}$ is an antilinear antiautomorphism of four order provided that $(v_{\alpha}^*)^*=-v_{\alpha}$ which define pseudoreal condition (see \cite{ScNaRi1978,BeTo1981,IvLeZu2004}). 

Therefore this form is called \textit{pseudoreal} and in terms of the Cartesian generators (\ref{pr6})-(\ref{pr7}) it is denoted by $\mathfrak{osp}^{*}(1|\mathfrak{o}(3))$. In terms of the CW generators $H:=\imath I_{3},E_{\pm}:=\imath I_{1}\mp I_{1},v_{+}:=v_{1},v_{-}:=v_{2}$ this pseudoreal form $\mathfrak{osp}^{*}(1|\mathfrak{o}(3))$ denoted also by $\mathfrak{osp}^{*}(1|\mathfrak{su}(2))$, is given as follows
\begin{eqnarray}\label{pr10}
&H^{*}\,=\,H,\quad E_{\pm}^{*}\,=\,E_{\mp},\quad v_{\pm}^{*}\,=\,\pm\varepsilon v_{\mp}.
\end{eqnarray} 

II. \textit{The noncompact real form $\mathfrak{osp}^{\dag}(1|\mathfrak{o}(2,1))\simeq\mathfrak{osp}^{\dag}(1|\mathfrak{sl}(2;\mathbb{R}))\simeq\mathfrak{osp}^{\dag}(1|\mathfrak{su}(1,1))$}.\\ 
In therms of the generators $\{I_{i},v_{\alpha}|i=1,2,3;\alpha=1,2\}$ this form is defined by the following conjugation:
\begin{eqnarray}\label{pr11}
&&I^{\dag}_{i}=(-1)^{i-1}I_{i},\;\; v^{\dag}_{1}=\varepsilon v_{1}\;\; v^{\dag}_{2}=\varepsilon v_{2},
\end{eqnarray}
where $\varepsilon=1$ if ($^{\dag}$) is the graded conjugation, and $\varepsilon=\imath$ if ($^{\dag}$) is the ungraded conjugation.\footnote{See (\ref{pr8}), (\ref{pr9}), where the conjugation ($^*$) is replaced by ($^{\dag}$).} We see that the conjugation ($^{\dag}$) prolonged to the odd generators $v_{\alpha}$ is an antilinear antiautomorphism of second order, that is $(v_{\alpha}^*)^*=v_{\alpha}$. Therefore this form is called \textit{real}, and it is denoted by $\mathfrak{osp}^{\dag}(1|\mathfrak{o}(2,1))$. 

If we introduce the CW generators 
\begin{eqnarray}\label{pr11'}
H:=\imath I_{3},\;\;E_{\pm}:=\imath I_{1}\mp I_{2},\;\;v_{+}:=v_{1},\;\;v_{-}:=v_{2}, 
\end{eqnarray}
where the Cartesian generators $\{I_{i},v_{\alpha}|i=1,2,3;\alpha=1,2\}$ satisfy the conjugation (\ref{pr11}), then the real condition is given as follows 
\begin{eqnarray}\label{pr12}
&&H^{\dag}=-H,\;\; E_{\pm}^{\dag}=-E_{\pm},\;\; v_{\pm}^{\dag}=\varepsilon v_{\pm}.
\end{eqnarray}\noindent
In terms of the given CW basis the real form $\mathfrak{osp}^{\dag}(1|\mathfrak{o}(2,1))$ is also denoted by $\mathfrak{osp}^{\dag}(1|\mathfrak{sl}(2;\mathbb{R}))$.

We can also introduce an alternative CW basis $H',E_{\pm}',v_{\pm}'$ in $\mathfrak{osp}^{\dag}(1|\mathfrak{o}(2,1))$ which are expressed in terms of the Cartesian generators $I_{i},v_{\alpha}\, (i=1,2,3;\alpha=1,2)$ and the CW generators $H,E_{\pm},v_{\pm}$ as follows:
\begin{eqnarray}
\begin{array}{rcl}\label{pr13}
&&H'=\imath I_{2}=\displaystyle-\frac{\imath}{2}\big(E_{+}-E_{-}\big),
\\[9pt]
&&E_{\pm}'=\imath I_{1}\pm I_{3}=\displaystyle\mp\imath H+\frac{1}{2}\big(E_{+}+E_{-}\big),
\\[9pt]
&&v_{+}'=\displaystyle\frac{1}{\sqrt{2}}\big(v_{+}+\imath v_{-}\big),\;\; v_{-}'=\displaystyle\frac{1}{\sqrt{2}}\big(\imath v_{+}+v_{-}\big).
\end{array}
\end{eqnarray}
The CW basis $H',E_{\pm}',v_{\pm}'$  satisfy the defining relations (\ref{pr1})--(\ref{pr3}), and it has the conjugation properties:
\begin{eqnarray}\label{pr14}
&&(H')^{\dag}=H',\;\;(E_{\pm}')^{\dag}=-E_{\mp}',\;\;(v_{\pm}')^{\dag}=-\imath\varepsilon v_{\mp}'.
\end{eqnarray}
The real superalgebra $\mathfrak{osp}^{\dag}(1|\mathfrak{o}(2,1))$ in terms of the CW basis $H',E_{\pm}',v_{\pm}'$ will be also denoted by $\mathfrak{osp}^{\dag}(1|\mathfrak{su}(1,1))$.

It should be noted that in the case of $\mathfrak{osp}^{\dag}(1|\mathfrak{su}(1,1))$ the Cartan generator $H'$ is compact while for the case $\mathfrak{osp}^{\dag}(1|\mathfrak{sl}(2,\mathbb{R}))$ the Cartan generator $H$ is noncompact.

It should be noted also that the Casimir element of two order, that is a $\mathfrak{osp}(1|2;\mathbb{C})$-invariant element of the universal enveloping superalgebra $U(\mathfrak{osp}(1|2;\mathbb{C}))$:
\begin{eqnarray}\label{pr15}
\begin{array}{rcl}
&&C_{2}:=\;\displaystyle v_{+}v_{-}-v_{-}v_{+}+\frac{1}{2}E_{+}E_{-}+\frac{1}{2}E_{-}E_{+}+H^2
\\[7pt]
&&\phantom{C_{2}:}=\;\displaystyle 2v_{+}v_{-}+E_{+}E_{-}+H^2-\frac{1}{2}H
\end{array}
\end{eqnarray}
satisfy the reality condition
\begin{eqnarray}\label{pr16}
&&C_{2}^{\divideontimes}\;=\;C_{2}^{}\quad (\divideontimes=*,\dag) 
\end{eqnarray}
with respect to all conjugations (\ref{pr10}), (\ref{pr12}) and (\ref{pr14}) provided that these conjugations act on the product of two homogeneous elements $x_a$ and $y_b$ by the formulas (\ref{pr8}), (\ref{pr9}), where the superbracket $[\![x_a,x_b]\!]$ should be replaced on the usual product $x_ax_b$.  

\setcounter{equation}{0}
\section{Classical $r$-matrices of the complex Lie superalgebra $\mathfrak{osp}(1|2;\mathbb{C})$} 
In this section we obtain complete classification bialgebras (classical $r$-matrices) for the complex Lie superalgebra $\mathfrak{osp}(1|2;\mathbb{C})$  and its real forms $\mathfrak{osp}^{*}(1|\mathfrak{o}(3))$ and $\mathfrak{osp}^{\dag}(1|\mathfrak{o}(2,1))$ using the isomorphisms: $\mathfrak{osp}(1|\mathfrak{o}(3;\mathbb{C}))\simeq\mathfrak{osp}(1|\mathfrak{sl}(2;\mathbb{C}))$, $\mathfrak{osp}^{*}(1|\mathfrak{o}(3))\simeq\mathfrak{osp}^{*}(1|\mathfrak{su}(2))$, $\mathfrak{osp}^{\dag}(1|\mathfrak{o}(2,1))\simeq\mathfrak{osp}^{\dag}(1|\mathfrak{sl}(2;\mathbb{R}))\simeq\mathfrak{osp}^{\dag}(1|\mathfrak{su}(1,1))$. In particular, we explicitly find out an isomorphism between $\mathfrak{osp}^{\dag}(1|\mathfrak{su}(1,1))$ and $\mathfrak{osp}^{\dag}(1|\mathfrak{sl}(2;\mathbb{R}))$ bialgebras and fix on the basis $\mathfrak{osp}^{\dag}(1|\mathfrak{o}(2,1))$ bialgebra in such forms which are convenient for quantizations.

By the definition any classical $r$-matrix of arbitrary complex or real Lie superalgebra $\mathfrak{g}=\mathfrak{g}_{\bar{0}}\oplus\mathfrak{g}_{\bar{1}}$, $r\in\mathfrak{g}_{\bar{0}}\wedge\mathfrak{g}_{\bar{0}}\oplus\mathfrak{g}_{\bar{1}}\wedge\mathfrak{g}_{\bar{1}}$, satisfy the classical Yang-Baxter equation (CYBE):
\begin{eqnarray}\label{rm1}
&&[r,\,r]_{S}=\tilde{\Omega}~.
\end{eqnarray}
Here $[\cdot,\cdot]_{S}$ is the graded Schouten bracket which for any monomial skew-symmetric even two-tensors $r_{1}^{}=x_{a}\wedge y_{a}$ and $r_{2}^{}=u_{b}\wedge v_{b}$ ($x_{a},y_{a}\in\mathfrak{g}_{a}$; $u_{b},v_{b}\in\mathfrak{g}_{b}$; $a,b,\in\{\bar{0},\bar{1}\})$ is given by
\begin{eqnarray}\label{rm2}
\begin{array}{rcl}
&&[x_{a}\wedge y_{a},\,u_{b}\wedge v_{b}]_{S}
\\[4pt]
&&\;\;:=x_{a}\wedge\bigl([\![y_{a},u_{b}]\!]\wedge{v_{b}}+(-1)^{ab}u_{b}\wedge[\![y_{a},v_{b}]\!]\bigr)
\\[4pt]
&&\;\;\;-(-1)^{a}y_{a}\wedge\bigl([\![x_{a},u_{b}]\!]\wedge{v_{b}}+(-1)^{ab}u_{b}\wedge[\![x_{a},v_{b}]\!]\bigr)
\\[4pt]
&&\;\;\;=[u_{b}\wedge v_{b},x_{a}\wedge y_{a}]_{S}
\end{array}
\end{eqnarray}
and $\tilde{\Omega}$ is the $\mathfrak{g}$-invariant element, $\tilde{\Omega}\in(\stackrel{3}\wedge\mathfrak{g})_{\mathfrak{g}}$, that in the case of $\mathfrak{g}:=\mathfrak{osp}(1|2;\mathbb{C})$ looks as follows:
\begin{eqnarray}\label{rm3}
\begin{array}{rcl}
&&\tilde{\Omega}={\gamma}\Omega(\mathfrak{osp}(1|2;\mathbb{C}))
\\[5pt]
&&\phantom{\Omega}=\gamma(4{E}_{-}\wedge H\wedge E_{+}+4{v}_{-}\wedge v_{+}\wedge H
\\[5pt]
&&\phantom{\Omega=}+2{v}_{-}\wedge v_{-}\wedge E_{+}-2v_{+}\wedge v_{+}\wedge {E}_{-}),
\end{array}
\end{eqnarray}
where $\gamma\in\mathbb{C}$. 

We have already mentioned that a classical $r$-matrix $r$ is an even two-tensor, i.e.:
\begin{eqnarray}\label{rm4}
\begin{array}{rcl}
&&r\;\in\;V_{\bar{0}}:=\mathfrak{osp}_{\bar{0}}(1|2;\mathbb{C})\wedge\mathfrak{osp}_{\bar{0}}(1|2;\mathbb{C})
\\[3pt]
&&\phantom{r\;\in\;V_{\bar{0}}:}\oplus\mathfrak{osp}_{\bar{1}}(1|2;\mathbb{C})\wedge\mathfrak{osp}_{\bar{1}}(1|2;\mathbb{C}). 
\end{array}
\end{eqnarray}
As a basis in the linear space $V_{0}$ we can take the following two-tensors: 
\begin{eqnarray}\label{rm5}
&&\begin{array}{rcl}
&&r_{0}:=2v_{+}\wedge v_{-}+E_{+}\wedge E_{-},  
\\[3pt]
&&r_{\pm}:=\pm v_{\pm}\wedge v_{\pm}\pm E_{\pm}\wedge H,
\end{array}
\\[5pt]
&&\begin{array}{rcl}\label{rm6}
&&\bar{r}_{0}:=E_{+}\wedge E_{-},\;\;\bar{r}_{\pm}:=\pm E_{\pm}\wedge H.
\end{array}
\end{eqnarray}
The following propositions are valid:\\
\textit{(i) Any linear combination of the elements (\ref{rm5}) is a classical $r$-matrix, namely, if
\begin{eqnarray}\label{rm7}
&&r:=\beta_{+}r_{+}+\beta_{0}r_{0}+\beta_{-}r_{-} 
\end{eqnarray}
for $\forall\,\beta_{+},\beta_{0},\beta_{-}\in\mathbb{C}$, then we have 
\begin{eqnarray}\label{rm8}
\begin{array}{rcl}
&&[r,r]_{S}=(\beta_{0}^{2}+\beta_{+}\beta_{-})\Omega(\mathfrak{osp}(1|2;\mathbb{C}))
\\[4pt]
&&\phantom{[r,r]_{S}}\equiv\gamma\Omega(\mathfrak{osp}(1|2;\mathbb{C})).
\end{array}
\end{eqnarray}
(ii) Any linear combination of the elements $\bar{r}_{0},\bar{r}_{\pm}$:
\begin{eqnarray}\label{rm9}
\bar{r}\!\!&:=\!\!&{\beta}_{+}\bar{r}_{+}+{\beta}_{0}\bar{r}_{0}+{\beta}_{-}\bar{r}_{-} 
\end{eqnarray}
for ${\beta}_{0}^{2}+{\beta}_{+}{\beta}_{-}=0$ satisfies the homogeneous CYBE, i.e. $[\bar{r},\bar{r}]_{S}=0$.\\
(iii) Any classical $r$-matrix of $\mathfrak{osp}(1|2;\mathbb{C})$ is presented only in the form (\ref{rm7}) or (\ref{rm9}).}

Firstly we prove the proposition \textit{(i)}. Let (\ref{rm7}) be an arbitrary linear combination of the elements (\ref{rm5}). Because all basis elements (\ref{rm5}) are classical $r$-matrices, moreover $[r_{\pm}^{},r_{\pm}^{}]_{S}=0$, as well as the Schouten brackets of the elements $r_{\pm}^{}$ with $r_{0}^{}$ are also equal to zero, $[r_{\pm},r_{0}^{}]_{S}=0$, and we have 
\begin{eqnarray}\label{rm10}
\begin{array}{rcl}
&&[r,r]_{S}=2\beta_{+}^{}\beta_{-}^{}[r_{+}^{},r_{-}^{}]_{S}+\beta_{0}^{2}[r_{0}^{},r_{0}^{}]_{S}
\\[4pt]
&&\phantom{[r,r]_{S}}=(\beta_{0}^{2}+\beta_{+}^{}\beta_{-}^{})\Omega(\mathfrak{osp}(1|2;\mathbb{C}))
\\[4pt]
&&\phantom{[r,r]_{S}}\equiv\gamma\Omega(\mathfrak{osp}(1|2;\mathbb{C})).
\end{array}
\end{eqnarray}
Thus the arbitrary element (\ref{rm7}) is a classical $r$-matrix.

In the case of the proposition \textit{(ii)} we have $[\bar{r}_{\pm}^{},\bar{r}_{\pm}^{}]_{S}=0$,  $[\bar{r}_{\pm},\bar{r}_{0}^{}]_{S}=0$, and the Schouten bracket for the arbitrary vector (\ref{rm9}) looks  as follows: 
\begin{eqnarray}\label{rm11}
\begin{array}{rcl}
&&[\bar{r},\bar{r}]_{S}=2{\beta}_{+}^{}{\beta}_{-}^{}[\bar{r}_{+}^{},\bar{r}_{-}^{}]_{S}+{\beta}_{0}^{2}[\bar{r}_{0}^{},\bar{r}_{0}^{}]_{S}
\\[4pt]
&&\phantom{[\bar{r},\bar{r}]_{S}}=({\beta}_{0}^{2}+{\beta}_{+}^{}{\beta}_{-}^{})\Omega(\mathfrak{sl}(2;\mathbb{C}))
\\[4pt]
&&\phantom{[\bar{r},\bar{r}]_{S}}\equiv{\gamma}\,\Omega(\mathfrak{sl}(2;\mathbb{C})),
\end{array}
\end{eqnarray}
where $\Omega(\mathfrak{sl}(2;\mathbb{C})$ is the $\mathfrak{sl}(2;\mathbb{C})$-invariant element (the first term in the parenthesis on the right-hand side of (\ref{rm3})). Because $\Omega(\mathfrak{sl}(2;\mathbb{C}))\neq\tilde{\Omega}(\mathfrak{osp}(1|2;\mathbb{C}))$, therefore it should be ${\gamma}:={\beta}_{0}^{2}+{\beta}_{+}^{}{\beta}_{-}^{}=0$ so  the element $\bar{r}$ satisfies the homogeneous CYBE (\ref{rm8}). 

Finally we prove the proposition \textit{(iii)}. Let us consider a general two-tensor $r_{g}$  which is a sum of the two-tensors (\ref{rm7}) and (\ref{rm9}) 
\begin{eqnarray}\label{rm12}
\begin{array}{rcl}
&&r_{g}=r+\bar{r}=\beta_{+}r_{+}+\beta_{0}r_{0}+\beta_{-}r_{-}
\\[4pt]
&&\phantom{r_{g}=r+\bar{r}}+{\beta}_{+}'\bar{r}_{+}+{\beta}_{0}'\bar{r}_{0}+{\beta}_{-}'\bar{r}_{-}
\end{array}
\end{eqnarray}
with arbitrary coefficients $\beta_{\pm},\beta_{0},{\beta}_{\pm}',{\beta}_{0}'\in\mathbb{C}$. It is not too difficult to see that the two-tensor (\ref{rm12}) is a non-zero classical $r$-matrix of $\mathfrak{osp}(1|2;\mathbb{C})$ if and only if $|\beta_{+}|+|\beta_{0}|+|\beta_{-}|\neq0$ and ${\beta}_{+}'={\beta}_{0}'={\beta}_{-}'=0$ or $|{\beta}_{+}'|+|{\beta}_{0}'|+|{\beta}_{-}'|\neq0$ and $\beta_{+}=\beta_{0}=\beta_{-}=0$ provided that ${\beta'_0}^{2}+{\beta}_{+}'{\beta}_{-}'=0$. Indeed, let us calculate the Schouten bracket of $r_{g}$; 
\begin{eqnarray}\label{rm13}
\begin{array}{rcl}
&&[r_{g},r_{g}]_{S}=[r,r]_{S}+[\bar{r},\bar{r}]_{S}+2[r,\bar{r}]_{S}
\\[4pt]
&&\;\;=(\beta_{0}^{2}+\beta_{+}\beta_{-})\Omega(\mathfrak{osp}(1|2;\mathbb{C}))
\\[4pt]
&&\;\;+({\beta_0'}^{2}+{\beta}_{+}'{\beta}_{-}')\Omega(\mathfrak{sl}(2;\mathbb{C})) 
\\[4pt]
&&\;\;+2\beta_{+}{\beta}_{+}'[r_{+},\bar{r}_{+}]_{S}+2(\beta_{+}{\beta}_{0}'[r_{+},\bar{r}_{0}]_{S}+\beta_{0}{\beta}_{+}'[r_{0},\bar{r}_{+}]_{S})
\\[4pt]
&&\;\;+2\big(\beta_{+}{\beta}_{-}'[r_{+},\bar{r}_{-}]_{S}+\beta_{0}{\beta}_{0}'[r_{0},\bar{r}_{0}]_{S}+\beta_{-}{\beta}_{+}'[r_{-},\bar{r}_{+}]_{S}\big)
\\[4pt]
&&\;\;+2(\beta_{-}{\beta}_{0}'[r_{-},\bar{r}_{0}]_{S}+\beta_{0}{\beta}_{-}'[r_{0},\bar{r}_{-}]_{S})+2\beta_{-}{\beta}_{-}'[r_{-},\bar{r}_{-}]_{S},
\end{array}
\end{eqnarray}
where all the Schouten brackets $[r_{i},\bar{r}_{j}]_{S}$, ($i,j=\pm,0$), are different\footnote{Moreover, they are linearly independent elements.} and each of them has the weight $(i1+j1)$ with respect to the $\mathop{\rm ad}_{H}$-action, $\mathop{\rm ad}_{H}([r_{i},\bar{r}_{j}]_{S})\equiv [H,[r_{i},\bar{r}_{j}]_{S}]_{S}=(i1+j1)[r_{i},\bar{r}_{j}]_{S}$, for example, the first and the last Schouten brackets $[r_{\pm},\bar{r}_{\pm}]_{S}$ have the weight $\pm2$. The two-tensor (\ref{rm12}) will be a classical $r$-matrix of $\mathfrak{osp}(1|2;\mathbb{C})$ if its Schouten bracket (\ref{rm13}) is $\mathfrak{osp}(1|2;\mathbb{C})$-invariant, i.e. 
\begin{eqnarray}\label{rm14}
[x,[r_{g},r_{g}]_{S}]_{S}= 0\;\;(\forall x\in\mathfrak{osp}(1|2;\mathbb{C})).
\end{eqnarray}
Applying this condition to the right-hand side of (\ref{rm13}) for $x=H$ first and then for $x=E_{+}$, and finally for $x=v_{+}$ we obtain the following quadratic equations: 
\begin{eqnarray}\label{rm15}
\begin{array}{rcl}
&&\beta_{\pm}{\beta}_{\pm}'=0,\;\;\beta_{\pm}{\beta}_{0}'=0,\;\;\beta_{0}{\beta}_{\pm}'=0,
\\[4pt]
&&2\beta_{\pm}{\beta}_{\mp}'+\beta_{0}{\beta}_{0}'=0,\quad{\beta_0'}^{2}+{\beta}_{+}'{\beta}_{-}'=0.
\end{array}
\end{eqnarray}
It is easy to see that if $\beta_{+}\neq0$ then ${\beta}_{+}'={\beta}_{-}'={\beta}_{0}'=0$. The other two cases $\beta_{-}\neq0$ and $\beta_{0}\neq0$ are similar. Thus the general two-tensor (\ref{rm12}), where at least one of the coefficients $\beta_{\pm}$, $\beta_{0}$ is not zero, will be  a classical $r$-matrix if it has the form (\ref{rm7}). It is evident that a similar result is obtained if we replace $\beta_{i}\rightleftarrows{\beta}_{i}'$ ($i=\pm,0$) provided that ${\beta_0'}^{2}+{\beta}_{+}'{\beta}_{-}'=0$, that is the general two-tensor (\ref{rm12}), where at least one of the coefficients ${\beta}_{\pm}'$, ${\beta}_{0}'$ is not zero, will be  a classical $r$-matrix if it has the form (\ref{rm9}). 

We shall call the parameter $\gamma=\beta_{0}^{2}+\beta_{+}^{}\beta_{-}^{}$ in (\ref{rm8}) the $\gamma$-characteristic of the classical $r$-matrix (\ref{rm7}). It is evident that the $\gamma$-characteristic of the classical $r$-matrix $r$ is invariant under the $\mathfrak{ospl}(1|2;\mathbb{C})$-automorphisms, i.e. any two $r$-matrices $r$ and $r'$, which are connected by a $\mathfrak{osp}(1|2;\mathbb{C})$-automorphism, have the same $\gamma$-characteristic, $\gamma=\gamma'$. 

There are two types of explicit {$\mathfrak{osp}(1|2;\mathbb{C})$}-automorphisms. 
First type connecting the classical $r$-matrices with zero $\gamma$-characteristic is given by the formulas:
\begin{eqnarray}\label{rm16}
\begin{array}{rcl}
&&\varphi_{0}(E_{+})=\chi(\tilde{\beta}_{+}E_{+}-2\tilde{\beta}_{0}\,H+\tilde{\beta}_{-}E_{-}),
\\[4pt]
&&\varphi_{0}(E_{-})=\chi^{-1}(\tilde{\beta}_{-}E_{+}-2\kappa\tilde{\beta}_{0}\,H+\tilde{\beta}_{+}E_{-}),
\\[4pt]
&&\varphi_{0}(H)=\tilde{\beta}_{0}\,E_{+}+(\kappa\tilde{\beta}_{+}+\tilde{\beta}_{-})\,H+\kappa\tilde{\beta}_{0}\,E_{-},
\\[4pt]
&&\varphi_{0}(v_{+})=\sqrt{\chi}\Big(\sqrt{\tilde{\beta}_{+}}v_{+}+\sqrt{\tilde{\beta}_{-}}v_{-}\Big),
\\[4pt]
&&\varphi_{0}(v_{-})=\sqrt{\chi^{-1}}\Big(\sqrt{\tilde{\beta}_{-}}v_{+}+\kappa\sqrt{\tilde{\beta}_{+}}v_{-}\Big),
\end{array}
\end{eqnarray}
where $\chi$ is a non-zero rescaling parameter (including $\chi=1$), $\kappa$ takes two values $+1$ or $-1$, and the parameters $\tilde{\beta}_{i}$ ($i=+,0,-$) satisfy the conditions:
\begin{eqnarray}\label{rm17}
&&\gamma:=\tilde{\beta}_{0}^{2}+\tilde{\beta}_{+}^{}\tilde{\beta}_{-}^{}=0,\;\;\kappa\tilde{\beta}_{+}^{}-\tilde{\beta}_{-}^{}=1.
\end{eqnarray}
Let us consider two independent pair $\{r,r'\}$ and $\{\bar{r},\bar{r}'\}$ of the general $r$-matrices with zero $\gamma$-characteristics: 
\begin{eqnarray}\label{rm18}
\begin{array}{rcl}
&&r:=\beta_{+}r_{+}+\beta_{0}r_{0}+\beta_{-}r_{-},
\\[4pt]
&&r':=\beta_{+}'r_{+}+\beta_{0}'r_{0}+\beta_{-}'r_{-},
\end{array}
\\[5pt]
\begin{array}{rcl}\label{rm19}
&&\bar{r}:=\beta_{+}\bar{r}_{+}+\beta_{0}\bar{r}_{0}+\beta_{-}\bar{r}_{-},
\\[4pt]
&&\bar{r}':=\beta_{+}'\bar{r}_{+}+\beta_{0}'\bar{r}_{0}+\beta_{-}'\bar{r}_{-}, 
\end{array}
\end{eqnarray} 
where $\gamma=\beta_{0}^{2}+\beta_{+}\beta_{-}=0$ and $\gamma=\beta_{0}^{'2}+\beta_{+}'\beta_{-}'=0$.  Moreover, we suppose that the parameters $\beta_{\pm}$ and $\beta_{\pm}'$ satisfy the additional relations:
\begin{eqnarray}\label{rm20}
\kappa\beta_{+}^{}-\beta_{-}^{}=\chi\beta_{+}'-\chi^{-1}\kappa\beta_{-}'\neq 0,
\end{eqnarray}
where the parameters $\kappa$ and $\chi$ are the same as in (\ref{rm16}).

One can check that the following formula are valid:
\begin{eqnarray}\label{rm21}
&&r=(\varphi_{0}\otimes\varphi_{0})r',
\\[3pt]\label{rm22}
&&\bar{r}=(\varphi_{0}\otimes\varphi_{0})\bar{r}',
\end{eqnarray}
where $\varphi_{0}$ is the $\mathfrak{osp}(1|2;\mathbb{C})$-automorphism (\ref{rm16}) with the following parameters:
\begin{eqnarray}\label{rm23}
\begin{array}{rcl}
&&\tilde{\beta}_{0}^{}=\displaystyle\frac{\beta_{0}^{}(\chi\beta_{+}'+\chi^{-1}\kappa\beta_{-}')-\beta_{0}'(\kappa\beta_{+}^{}+\beta_{-})}{(\kappa\beta_{+}^{}-\beta_{-}^{})(\chi\beta_{+}'-\chi^{-1}\kappa\beta_{-}')},
\\[11pt]
&&\tilde{\beta}_{+}^{}=\displaystyle\frac{\kappa(\kappa\beta_{+}^{}+\beta_{-}^{})(\chi\beta_{+}'+\chi^{-1}\kappa\beta_{-}')+4\beta_{0}^{}\beta_{0}'}{2(\kappa\beta_{+}^{}-\beta_{-}^{})(\chi\beta_{+}'-\chi^{-1}\kappa\beta_{-}')}+\frac{\kappa}{2},
\\[11pt]
&&\tilde{\beta}_{-}^{}=\displaystyle\frac{(\kappa\beta_{+}^{}+\beta_{-}^{})(\chi\beta_{+}'+\chi^{-1}\kappa\beta_{-}')+4\kappa\beta_{0}^{}\beta_{0}'}{2(\kappa\beta_{+}^{}-\beta_{-}^{})(\chi\beta_{+}'-\chi^{-1}\kappa\beta_{-}')}-\frac{1}{2}.
\end{array}
\end{eqnarray}
It is easy to check that the formulas (\ref{rm23}) satisfy the condition $\tilde{\beta}_{0}^{2}+\tilde{\beta}_{+}\tilde{\beta}_{-}=0$. 

Let us assume in (\ref{rm23}) that the parameters $\beta_{0}'$ and $\beta_{-}'$ are equal to zero. Then the general classical $r$-matrix $r$, satisfying the homogeneous CYBE, is reduced to usual Jordanian form by the automorphism $\varphi_{0}^{}$ with the parameters: 
\begin{eqnarray}\label{rm24}
\tilde{\beta}_{0}^{}=\frac{\beta_{0}^{}}{\kappa\beta_{+}^{}-\beta_{-}^{}},\;\;   
\tilde{\beta}_{\pm}^{}=\frac{\beta_{\pm}^{}}{\kappa\beta_{+}^{}-\beta_{-}^{}}.
\end{eqnarray}

Second type of $\mathfrak{osp}(1|2;\mathbb{C})$-automorphism connecting the classical $r$-matrices with non-zero $\gamma$-characteristic is given as follows
\begin{eqnarray}\label{rm25}
\begin{array}{rcl}
&&\varphi_{1}^{}(E_{+})=\displaystyle\frac{\chi}{2}\Bigl((\tilde{\beta}_{0}+1)\,E_{+}+2\tilde{\beta}_{-}H-\frac{\tilde{\beta}_{-}^{2}}{\tilde{\beta}_{0}+1}E_{-}\Bigr),
\\[8pt]
&&\varphi_{1}^{}(E_{-})=\displaystyle\frac{\chi^{-1}}{2}\Bigl(\frac{-\tilde{\beta}_{+}^{2}}{\tilde{\beta}_{0}+1}E_{+}+2\tilde{\beta}_{+}H+(\tilde{\beta}_{0}+1)E_{-}\Bigr),
\\[8pt]
&&\varphi_{1}^{}(H)=\displaystyle\frac{1}{2}\bigl(-\tilde{\beta}_{+}E_{+}+2\tilde{\beta}_{0}H-\tilde{\beta}_{-}E_{-}\bigr),
\\[8pt]
&&\varphi_{1}(v_{+})=\displaystyle\sqrt{\frac{\chi}{2}}\Big(\sqrt{\tilde{\beta}_{0}+1}v_{+}+\frac{\tilde{\beta}_{-}}{\sqrt{\beta_{0}+1}}v_{-}\Big),
\\[10pt]
&&\varphi_{1}(v_{-})=\displaystyle\sqrt{\frac{\chi^{-1}}{2}}\Big(\frac{\tilde{\beta}_{+}}{\sqrt{\beta_{0}+1}}v_{+}+\sqrt{\tilde{\beta}_{0}+1}v_{-}\Big),
\end{array}
\end{eqnarray}
where $\chi$ is a non-zero rescaling parameter, and $\tilde{\beta}_{0}^{2}+\tilde{\beta}_{+}\tilde{\beta}_{-}=1$.

Let us consider two general $r$-matrices with non-zero $\gamma$-characteristics:
\begin{eqnarray}\label{rm26}
\begin{array}{rcl}
&&r:=\beta_{+}r_{+}+\beta_{0}r_{0}+\beta_{-}r_{-},
\\[4pt]
&&r':=\beta_{+}'r_{+}+\beta_{0}'r_{0}+\beta_{-}'r_{-}, 
\end{array}
\end{eqnarray}
where the parameters $\beta_{\pm}$, $\beta_{0}$ and $\beta_{\pm}'$, $\beta_{0}'$ can be equal to zero provided that $\gamma=\beta_{0}^2+\beta_{+}\beta_{-}=\gamma'= (\beta_{0}')^2+\beta_{+}'\beta_{-}'\neq0$, i.e. both $r$-matrices $r$ and $r'$ have the same non-zero $\gamma$-characteristic $\gamma=\gamma'\neq0$. 

One can check the following relation: 
\begin{eqnarray}\label{rm27}
&&r=(\varphi_{1}^{}\otimes\varphi_{1}^{})r',
\end{eqnarray}
where $\varphi_{1}$ is the $\mathfrak{osp}(1|2;\mathbb{C})$-automorphism (\ref{rm25}) with the parameters:
\begin{eqnarray}\label{rm28}
\begin{array}{rcl}
&&\tilde{\beta}_{0}^{}=\displaystyle\frac{(\beta_{0}^{}+\beta_{0}')^{2}-(\beta_{+}^{}-\chi\beta_{+}')(\beta_{-}^{}-\chi^{-1}\beta_{-}')}{(\beta_{0}^{}+\beta_{0}')^{2}+(\beta_{+}-\chi\beta_{+}')(\beta_{-}-\chi^{-1}\beta_{-}')},
\\[10pt]
&&\tilde{\beta}_{\pm}^{}=\displaystyle\frac{2(\beta_{0}^{}+\beta_{0}')(\beta_{\pm}^{}-\chi^{\pm1}\beta_{\pm}')}{(\beta_{0}^{}+\beta_{0}')^{2}+(\beta_{+}^{}-\chi\beta_{+}')(\beta_{-}^{}-\chi^{-1}\beta_{-}')}.
\end{array}
\end{eqnarray}
It is easy to check that the formulas (\ref{rm28}) satisfy the condition $\tilde{\beta}_{0}^2+\tilde{\beta}_{+}^{}\tilde{\beta}_{-}^{}=1$.

If we assume in  (\ref{rm26}) that the parameters $\beta_{\pm}'$ are equal to zero then the general classical $r$-matrix $r$, satisfying the non-homogeneous CYBE, is reduced to the usual standard form by the automorphism $\varphi_{1}$, (\ref{rm25}), with the following parameters:
\begin{eqnarray}\label{rm29}\tilde{\beta}_{0}^{}=\frac{\beta_{0}^{}}{\beta_{0}'},\;\;   
\tilde{\beta}_{\pm}^{}=\frac{\beta_{\pm}^{}}{\beta_{0}'}.
\end{eqnarray} 
Finally for $\mathfrak{osp}(1|2,\mathbb{C})$ we get the following result:

\textit{For the complex Lie superalgebra $\mathfrak{osp}(1|2,\mathbb{C})$ there exists up to $\mathfrak{osp}(1|2,\mathbb{C})$ automorphisms three solutions of CYBE, namely Jordanian $r_{J}$, super-Jordanian $r_{sJ}$ and standard $r_{st}$:
\begin{eqnarray}\label{rm30}
&&\begin{array}{rcl}
&&r_{J}=\beta E_{+}\wedge H,\;\;[r_{J},r_{J}]_{S}=0,
\end{array}
\\[5pt] 
&&\begin{array}{rcl}\label{rm31} 
&&r_{sJ}=\beta_{1}(E_{+}\wedge H+v_{+}\wedge v_{+}),
\\[3pt] 
&&[r_{sJ},r_{sJ}]_{S}=0,
\end{array}
\\[5pt] 
&&\begin{array}{rcl}\label{rm32}
&&r_{st}=\beta_{0}(E_{+}\wedge E_{-}+2v_{+}\wedge v_{-}),
\\[3pt] 
&&[r_{st},r_{st}]_{S}=\beta_{0}^{2}\Omega,
\end{array}
\end{eqnarray}
where the complex parameters $\beta$ and $\beta_{1}$ can be removed by the rescaling $\mathfrak{osp}(1|2,\mathbb{C})$-automorphism: $\varphi(E_{+})=\beta^{-1}E_{+}$, $\varphi(E_{-})=\beta E_{-}$, $\varphi(v_{+})=\sqrt{\beta^{-1}}\,v_{+}$, $\varphi(v_{-})=\sqrt{\beta}\,v_{-}$, $\varphi(H)=H$; the parameter $\beta_{0}=e^{\imath\phi}|\beta_{0}|$ for $|\phi|\leq\frac{\pi}{2}$ is effective.}

So, we obtained the full classification of all nonequivalent quasitriangular Lie bisuperalgebras ($\mathfrak{osp}(1|2,\mathbb{C}),\delta_{(r)}$) in terms of all nonequivalent (nonisomorphic) classical $r$-matrices (\ref{rm30})-(\ref{rm32}). It should be noted that the classical $r$-matrices for the complex Lie superalgebra $\mathfrak{osp}(1|2,\mathbb{C})$ was obtained previously in \cite{JuSo1998} where the authors used a technique of computer algebraic computations. 

\setcounter{equation}{0}
\section{Classical $r$-matrices of $\mathfrak{osp}(1|2;\mathbb{C})$ real forms} 
The coproduct $\Delta_0$ (\ref{in7}), which is a homomorphism $\mathfrak{g}\stackrel{\Delta_0}{\longrightarrow}\mathfrak{g}\otimes\mathfrak{g}$ for any complex Lie superalgebra $\mathfrak{g}$ (and $\mathfrak{g}=\mathfrak{osp}(1|2;\mathbb{C})$ particularly), can be induced on all the universal enveloping superalgebra $U(\mathfrak{g})\stackrel{\Delta_0}{\longrightarrow}U(\mathfrak{g})\otimes U(\mathfrak{g})$ using the linearity and multiplicativity:
\begin{eqnarray}\label{rmr1}
\begin{array}{rcl}
&&\Delta_0(X+Y)=\Delta_0(X)+\Delta_0(Y),
\\[4pt]
&&\Delta_0(XY)=\Delta_0(X)\Delta_0(Y)
\end{array}
\end{eqnarray}
for $\forall\, X,Y\in U(\mathfrak{g})$. These relations can be used to lift the $\divideontimes$-conjugation, ($\divideontimes=*,\dag$), from $U(\mathfrak{g})$ to the tensor product $U(\mathfrak{g})\otimes U(\mathfrak{g})$. It should be noted that the coproduct $\Delta_0$ and the $\divideontimes$-conjugation preserve the $\mathbb{Z}_2$-grading $\deg(\cdot)$: $\deg(\Delta_0(X_a))=\deg(X_a^{\divideontimes})=\deg(X_a)=a\in\{\bar{0},\bar{1}\}$ for any homogeneous element $X_a\in U(\mathfrak{g})=U(\mathfrak{g})_{\bar{0}}\oplus U(\mathfrak{g})_{\bar{1}}$. There are two reality conditions for the coproduct $\Delta_0$:
\begin{eqnarray}\label{rmr2}
\begin{array}{rcl}
&&\Delta_0((X_a)^{\divideontimes})=(\Delta_0(X_a))^{\divideontimes\otimes\divideontimes},
\end{array}
\\[7pt]
\begin{array}{rcl}\label{rmr3}
&&\Delta_0((X_a)^{\divideontimes})=(\Delta_0(X_a))^{\divideontimes\tilde{\otimes}\divideontimes}
\\[3pt]
&&\phantom{a}
:=(\tau\Delta_0(X_a))^{\divideontimes{\otimes}\divideontimes}=\tau(\Delta_0(X_a))^{\divideontimes{\otimes}\divideontimes}
\end{array}
\end{eqnarray}

 for any homogeneous element $X_a\in U(\mathfrak{g})$, $a\in\{\bar{0},\bar{1}\}$. Here $\tau$ is a \textit{superpermutation} linear operator (\textit{the superflip}) in $U(\mathfrak{g})\otimes U(\mathfrak{g})$:  
\begin{eqnarray}\label{rmr4}
&&\tau(X_a\otimes Y_b)=(-1)^{ab}(Y_b\otimes X_a)
\end{eqnarray}
for all homogeneous elements $X_a,Y_b\in U(\mathfrak{g})$, $a,b\in\{\bar{0},\bar{1}\}$. We note also that the definition of the Hopf superalgebra differs from that of the usual Hopf algebra by the \textit{supermultiplication} of tensor product:
\begin{eqnarray}\label{rmr5}
&&(X_a\otimes Y_b)(V_c\otimes W_d)=(-1)^{bc}(X_aV_c\otimes Y_bW_d)
\end{eqnarray}
for all homogeneous elements $X_a,Y_b,V_c,W_d\in U(\mathfrak{g})$, $a,b,c,d\in\{\bar{0},\bar{1}\}$. The conjugation $\divideontimes\otimes\divideontimes$ in (\ref{rmr2}) (correspondingly $\divideontimes\tilde{\otimes}\divideontimes$ in (\ref{rmr3})) will be named \textit{direct} (correspondingly \textit{superflipped}) one.

From the reality condition (\ref{rmr2}) we find that: 
\begin{eqnarray}\label{rmr6}
\begin{array}{rcl}
&&{if}\;\;(X_aY_b)^{\divideontimes}=(-1)^{ab}Y_b^{\divideontimes}X_a^{\divideontimes}\;\;
({graded})
\;\;\textit{then}
\\[5pt]
&&(X_a\otimes Y_b)^{\divideontimes\otimes\divideontimes}=(X_a^{\divideontimes}\otimes Y_b^{\divideontimes})\;\;({ ungraded}),
\end{array}
\end{eqnarray}
and: 
\begin{eqnarray}\label{rmr7}
\begin{array}{rcl}
&&{if}\;\;(X_aY_b)^{\divideontimes}=Y_b^{\divideontimes}X_a^{\divideontimes}\;\;({ungraded})\;\;{then}\;\;
\\[5pt]
&&(X_a\otimes Y_b)^{\divideontimes\otimes\divideontimes}=(-1)^{ab}(X_a^{\divideontimes}\otimes Y_b^{\divideontimes})\;\;
({graded}).
\end{array}
\end{eqnarray}
From the reality condition (\ref{rmr3}) we find that: 
\begin{eqnarray}\label{rmr8}
\begin{array}{rcl}
&&{if}\;\;(X_aY_b)^{\divideontimes}=(-1)^{ab}Y_b^{\divideontimes}X_a^{\divideontimes}\;\;({graded})\;\;{then}\;\;
\\[5pt]
&&(X_a\otimes Y_b)^{\divideontimes\tilde{\otimes}\divideontimes}=(-1)^{ab}(Y_b^{\divideontimes}\otimes X_a^{\divideontimes})\;\;({graded}),
\end{array}
\end{eqnarray}
and: 
\begin{eqnarray}\label{rmr9}
\begin{array}{rcl}
&&{if}\;\;(X_aY_b)^{\divideontimes}=Y_b^{\divideontimes}X_a^{\divideontimes}\;\;({ungraded})\;\;{then}\;\;
\\[5pt]
&&(X_a\otimes Y_b)^{\divideontimes\tilde{\otimes}\divideontimes}=(Y_b^{\divideontimes}\otimes X_a^{\divideontimes})\;\;({ungraded}).
\end{array}
\end{eqnarray}

Next, we will use only the direct conjugation type (\ref{rmr2}), (\ref{rmr6}), (\ref{rmr7}) in the tensor product $U(\mathfrak{g})\otimes U(\mathfrak{g})$ since the results for the cases (\ref{rmr8}), (\ref{rmr9}) can be obtained from the cases (\ref{rmr6}), (\ref{rmr7}) using the superflip (\ref{rmr4}) explicitly.

The general non-reduced expression of the classical $r$-matrix (\ref{rm7}) (and also (\ref{rm9})) is convenient for the application of reality conditions:  
\begin{eqnarray}\label{rmr10}
&&r^{\divideontimes\otimes\divideontimes}=\beta_{+}^{*}r_{+}^{\divideontimes\otimes\divideontimes}+\beta_{0}^{*}r_{0}^{\divideontimes\otimes\divideontimes}+\beta_{-}^{*}r_{-}^{\divideontimes\otimes\divideontimes}=-r,
\end{eqnarray}
where $\divideontimes$ is the conjugation associated with corresponding real form ($\divideontimes=*,\dag$), and $\beta_{i}^{*}$ ($i=+,0,-$) means the complex conjugation of the number $\beta_{i}^{}$.  Moreover, if $r$-matrix is \textit{anti-real} (anti-Hermitian), i.e. it satisfies the condition (\ref{rmr10}), then its $\gamma$-characteristic is real. Indeed, applying the conjugation $\divideontimes$ to CYBE (\ref{rm8}) we have for the left-side:
\begin{eqnarray}\label{rmr11}
&&[r,r]_{S}^{\divideontimes\otimes\divideontimes\otimes\divideontimes}=-[r^{\divideontimes\otimes\divideontimes},r^{\divideontimes\otimes\divideontimes}]_{S}=-[r,r]_{S}=-\gamma\Omega 
\end{eqnarray}
and for the right-side:
\begin{eqnarray}\label{rmr12}
&&(\gamma\Omega)^{\divideontimes\otimes\divideontimes\otimes\divideontimes}=-\gamma^{*}\Omega 
\end{eqnarray}
for all real forms $\mathfrak{osp}^{*}(1|\mathfrak{su}(2))$, $\mathfrak{osp}^{\dag}(1|\mathfrak{su}(1,1))$, $\mathfrak{osp}^{\dag}(1|\mathfrak{su}(2;\mathbb{R}))$. It follows that the parameter $\gamma$ is real, $\gamma^{*}=\gamma$. 

I. {\large\textit{The compact pseudoreal real form} $\mathfrak{osp}^{*}(1|\mathfrak{su}(2))$}
($H^{*}=H$, $E_{\pm}^{*}=E_{\mp}$, $v_{\pm}^{*}=\pm\varepsilon v_{\mp}$).\\ 
It is not difficult to see that in the case of the grade ($\varepsilon=1$) and ungrade ($\varepsilon=\imath$) conjugations the basic two-tensors (\ref{rm5}) and (\ref{rm6}) have the following reality properties
\begin{eqnarray}\label{rmr13}
&&r_{0}^{*\otimes*}=-r_{0}^{},\quad r_{\pm}^{*\otimes*}=-r_{\mp}^{},
\\[4pt]\label{rmr14}
&&\bar{r}_{0}^{*\otimes*}=-\bar{r}_{0}^{},\quad\bar{r}_{\pm}^{*\otimes*}=-\bar{r}_{\mp}^{},
\end{eqnarray}
In this case it follows from (\ref{rmr10}) that
\begin{eqnarray}\label{rmr15}
\beta_{0}^{*}=\beta_{0}^{},\quad\beta_{\pm}^{*}=\beta_{\mp}^{}.
\end{eqnarray}
If in (\ref{rm7}) and (\ref{rm9}) $\gamma=\beta_{0}^{2}+\beta_{+}^{}\beta_{-}^{}=0$ then under conditions (\ref{rmr15}) we have $\beta_{0}^{}\beta_{0}^{*}+\beta_{\pm}^{}\beta_{\pm}^{*}=0$ and it follows that $\beta_{0}=\beta_{+}^{}=\beta_{-}^{}=0$, i.e. \textit{any classical $r$-matrix, which satisfies the homogeneous CYBE and the $\mathfrak{osp}^{*}(1|\mathfrak{su}(2))$ reality condition, is equal zero}. 

If in (\ref{rm7})  $\gamma=\beta_{0}^{2}+\beta_{+}^{}\beta_{-}^{}\neq0$  we have three $\mathfrak{osp}^{*}(1|\mathfrak{su}(2))$ real classical $r$-matrices: 
\begin{eqnarray}\label{rmr16}
\begin{array}{rcl}
&&r_{1}^{}=\beta_{0}^{}r_{0}^{},\;\;r_{2}^{}=\beta_{+}^{}r_{+}^{}+\beta_{+}^{*}r_{-}^{},
\\[4pt] 
&&r_{3}^{}=\beta_{+}'r_{+}^{}+\beta_{0}'r_{0}^{}+{\beta_{+}'}^{\!\!*}r_{-}^{},
\end{array}
\end{eqnarray} 
where $\beta_{0}^{}$ and $\beta_{0}'$ are real numbers and we use the conditions (\ref{rmr15}). 
The $r$-matrices $r_{i}$ ($i=1,2,3$) satisfy the non-homogeneous CYBE
\begin{eqnarray}\label{rmr17}
&&[r_{i}^{},r_{i}^{}]_S=\gamma_{i}\Omega,
\end{eqnarray}
where all $\gamma_{i}$ ($i=1,2,3$) are positive: $\gamma_{1}=\beta_{0}^{2}>0$, $\gamma_{2}=\beta_{+}^{}\beta_{+}^{*}>0$, $\gamma_{3}={\beta_{0}'}^{\!2}+\beta_{+}'{\beta_{+}'}^{\!\!*}>0$. 

Let two general classical $r$-matrices (\ref{rm26}) with non-zero $\gamma$-characteristic be $\mathfrak{osp}^{*}(1|\mathfrak{su}(2))$-antireal, i.e. their parameters satisfy the reality conditions (\ref{rmr15}). It follows that the functions (\ref{rm28}) for $\chi=e^{\imath\phi}$ have the same conjugation properties, i.e. $\tilde{\beta}_{0}^{*}=\tilde{\beta}_{0}^{}$, $\tilde{\beta}_{\pm}^{*}=\tilde{\beta}_{\mp}^{}$, and we obtain that the automorphism (\ref{rm24}) with such parameters is $\mathfrak{osp}^{*}(1|\mathfrak{su}(2))$-real, i.e.:  
\begin{eqnarray}\label{rmr18}
\begin{array}{rcl}
&&\varphi_{1}^{}(E_{\pm})^{*}=\varphi_{1}^{}(E_{\pm}^{*})=\varphi_{1}^{}(E_{\mp}),
\\[3pt]
&&\varphi_{1}^{}(H)^{*}=\varphi_{1}^{}(H^{*})=\varphi_{1}^{}(H),
\\[3pt]
&&\varphi_{1}^{}(v_{\pm})^{*}=\varphi_{1}^{}(v_{\pm}^{*})=\pm\varepsilon\varphi_{1}^{}(v_{\mp}).
\end{array}
\end{eqnarray}
We see that the $r$-matrices $r_{2}^{}$ and $r_{3}$ in (\ref{rmr16}) can be reduced to the standard $r$-matrix $r_{st}^{}:= r_{1}^{}$ using the formula (\ref{rm27}). 

It is easy to see that the standard $r$-matrix $r_{st}^{}=r_{1}^{}$ in (\ref{rmr16}) effectively depends only on positive values of the parameter $\alpha:=\beta_{0}^{}$. Indeed, we see that
\begin{eqnarray}\label{rmr19}
\begin{array}{rcl}
&&\alpha\bigl(\varphi(E_{+})\wedge\varphi(E_{-})+2\varphi(v_{+})\wedge\varphi(v_{-})\bigr)
\\[3pt]
&&\phantom{aaaaaaa}=-\alpha(E_{+}\wedge E_{-}+2v_{+}\wedge v_{-}),
\end{array}
\end{eqnarray}
where $\varphi$ is the simple $\mathfrak{osp}^{*}(1|\mathfrak{su}(2))$-pseudoreal automorphism: $\varphi(E_{\pm})=E_{\mp}$, $\varphi(H)=-H$, $\varphi(v_{\pm})=\imath v_{\mp}$, i.e. any negative value of parameter $\alpha$ in $r_{st}^{}$ can be replaced by the positive one.

We obtain the following result:\\
\textit{For the compact pseudoreal form $\mathfrak{osp}^{*}(1|\mathfrak{su}(2))$ with the graded ($\varepsilon=1$) or ungraded ($\varepsilon=\imath$) conjugation $({}^*)$, there exists up to the $\mathfrak{osp}^{*}(1|\mathfrak{su}(2))$-automorphisms only one solution of CYBE and this solution is the usual standard supersymmetric classical $r$-matrix $r_{st}$:
\begin{eqnarray}\label{rmr20}
\begin{array}{rcl}
&&r_{st}=\alpha(E_{+}\wedge E_{-}+2v_{+}\wedge v_{-}),
\\[3pt]
&&[r_{st},r_{st}]_S=\gamma\Omega,
\end{array}
\end{eqnarray}
where the effective parameter $\alpha$ is a positive number, and $\gamma=\alpha^{2}$.}
  
II. \textit{\large The non-compact real form} $\mathfrak{osp}^{\dag}(1|\mathfrak{sl}(2;\mathbb{R}))$ ($H^{\dag}=-H$, $E_{\pm}^{\dag}=-E_{\pm}$, $v_{\pm}^{\dag}=\varepsilon v_{\mp}$).\\
It is not difficult to see that in the case of the grade ($\varepsilon=1$) and ungrade ($\varepsilon=\imath$) conjugations the basic two-tensors (\ref{rm5}) and (\ref{rm6}) have the following reality properties
\begin{eqnarray}\label{rmr21}
&&r_{0}^{\dag\otimes\dag}=r_{0}^{},\quad r_{\pm}^{\dag\otimes\dag}=r_{\pm}^{},
\\[4pt]\label{rmr22}
&&\bar{r}_{0}^{\dag\otimes\dag}=\bar{r}_{0}^{},\quad \bar{r}_{\pm}^{\dag\otimes\dag}=\bar{r}_{\pm}^{},
\end{eqnarray}
In this case from (\ref{rmr10}) we obtain
\begin{eqnarray}\label{rmr23}
&&\beta_{0}^{*}=-\beta_{0},\quad\beta_{\pm}^{*}=-\beta_{\pm},
\end{eqnarray} 
i.e. all parameters $\beta_{i}^{}$ ($i=+,0,-$) are purely imaginary. 

Consider the general r-matrices (\ref{rm7}) and (\ref{rm9}) with the coefficients $\beta_{i}$ and $\bar{\beta}_{i}$ ($i=\pm,0$) satisfying the condition  $\beta_{0}^{2}+\beta_{+}^{}\beta_{-}^{}=0$ and (\ref{rmr23}) then we have six $\mathfrak{osp}^{\dag}(1|\mathfrak{su}(2;\mathbb{R}))$ solutions of the homogeneous CYBE :
\begin{eqnarray}\label{rmr24}
\begin{array}{rcl}
&&r_{1}=\beta_{+}^{}r_{+},\quad r_{2}=\beta_{-}^{}r_{-},
\\[3pt]
&&r_{3}=\beta_{+}'r_{+}+\beta_{0}'r_{0}+\beta_{-}'r_{-},
\end{array}
\\[5pt]
\begin{array}{rcl}\label{rmr25}
&&\bar{r}_{1}=\beta_{+}^{}\bar{r}_{+},\quad\bar{r}_{2}=\beta_{-}^{}\bar{r}_{-},
\\[3pt]
&&\bar{r}_{3}=\beta_{+}'\bar{r}_{+}+\beta_{0}'\bar{r}_{0}+\beta_{-}'\bar{r}_{-},
\end{array}
\end{eqnarray}
where all parameters $\beta_{i}^{}$ ($i=+,-$), $\beta_{i}'$ ($i=+,0,-$) are purely imaginary, and ${\beta_{0}'}^{\!2}+\beta_{+}'\beta_{-}'=0$. 

If the classical $r$-matrices (\ref{rm18}) and (\ref{rm19})  are $\mathfrak{osp}^{\dag}(1|\mathfrak{sl}(2;\mathbb{R}))$-antireal, i.e. their parameters satisfy the reality conditions (\ref{rmr23}), then for the real parameter $\chi$ all functions (\ref{rm23}) are real, i.e. $\tilde{\beta}_{0}^{*}=\tilde{\beta}_{0}^{}$, $\tilde{\beta}_{\pm}^{*}=\tilde{\beta}_{\pm}^{}$. We obtain that the automorphism of the type (\ref{rm16}) with such parameters is $\mathfrak{osp}^{\dag}(1|\mathfrak{sl}(2;\mathbb{R}))$-real, i.e.:  
\begin{eqnarray}\label{rmr26}
\begin{array}{rcl}
&&\varphi_{0}^{\dag}(E_{\pm})=\varphi_{0}^{}(E_{\pm}^{\dag})=-\varphi_{0}^{}(E_{\pm}),
\\[4pt]
&&\varphi_{0}^{\dag}(H)=\varphi_{0}^{}(H^{\dag})=-\varphi_{0}^{}(H),
\\[4pt]
&&\varphi_{0}^{\dag}(v_{\pm})=\varphi_{0}^{}(v_{\pm}^{\dag})=\varepsilon\varphi_{0}^{}(v_{\pm}).
\end{array}
\end{eqnarray}
It allows to reduce the $r$-matrices $r_{2}^{}$ and $r_{3}^{}$ in (\ref{rmr24}) to the  super-Jordanian $r$-matrix $r_{sJ}^{}:=r_{1}^{}$ and the $r$-matrices $\bar{r}_{2}^{}$ and $\bar{r}_{3}^{}$ in (\ref{rmr25}) to the  Jordanian $r$-matrix $r_{J}^{}:=\bar{r}_{1}^{}$ by using the formulas (\ref{rm21}) and  (\ref{rm22}).

In the case $\beta_{0}^{2}+\beta_{+}^{}\beta_{-}^{}\neq0$ in (\ref{rm7}) we have seven versions of $\mathfrak{osp}^{\dag}(1|\mathfrak{sl}(2;\mathbb{R}))$-real classical $r$-matrices. Five of them are with negative values of $\gamma_{i}$, ($i=1,2,\ldots,5$):
\begin{eqnarray}\label{rmr27}
\begin{array}{rcl}
&&r_{1}^{}=\beta_{0}^{}r_{0}^{},\qquad\qquad r_{2}^{}=\beta_{+}^{}r_{+}^{}+\beta_{0}^{}r_{0}^{},
\\[4pt]
&&r_{3}^{}=\beta_{0}^{}r_{0}^{}+\beta_{-}^{}r_{-}^{},
\;\;r_{4}^{}=\beta_{+}'r_{+}^{}+\beta_{-}'r_{-}^{},
\\[4pt]
&&r_{5}^{}=\beta_{+}''r_{+}^{}+\beta_{0}''r_{0}^{}+\beta_{-}''r_{-}^{},
\\[4pt]
&&[r_{i}^{},r_{i}^{}]_{S}=\gamma_{i}\Omega\qquad (i=1,2,\ldots,5),
\end{array}
\end{eqnarray}
where all parameters $\beta$ are purely imaginary, and $\gamma_{1}=\gamma_{2}=\gamma_{3}=\beta_{0}^{2}<0$, $\gamma_{4}=\beta_{+}'\beta_{-}'<0$, $\gamma_{5}=\beta_{0}''+\beta_{+}''\beta_{-}''<0$. The remaining two $r$-matrices $r_{i}^{}$ ($i=6,7$) have positive values of $\gamma_{i}$:
\begin{eqnarray}\label{rmr28}
\begin{array}{rcl}
&&r_{6}^{}=\beta_{+}'''r_{+}^{}+\beta_{-}'''r_{-}^{}, 
\\[4pt]
&&r_{7}^{}=\beta_{+}''''r_{+}^{}+\beta_{0}''''r_{0}^{}+\beta_{-}''''^{}r_{-}^{},
\\[4pt]
&&[r_{i}^{},r_{i}^{}]_{S}=\gamma_{i}\Omega'\qquad (i=6,7),
\end{array}
\end{eqnarray}
where $\gamma_{6}=\beta_{+}'''\beta_{-}'''>0$ and $\gamma_{7}={\beta_{0}''''}^{2}+\beta_{+}''''\beta_{-}''''>0$. 

Let two classical $r$-matrices (\ref{rm26}) be $\mathfrak{osp}^{\dag}(1|\mathfrak{sl}(2;\mathbb{R})$-real, i.e.  with their parameters satisfying the reality conditions (\ref{rmr23}). In such way the functions (\ref{rm28}) for real $\chi$ are real, i.e. $\tilde{\beta}_{0}^{*}=\tilde{\beta}_{0}^{}$, $\tilde{\beta}_{\pm}^{*}=\tilde{\beta}_{\pm}^{}$, and we obtain that the automorphism (\ref{rm25}) with such parameters is $\mathfrak{osp}^{\dag}(1|\mathfrak{sl}(2;\mathbb{R})$-real. We can conclude that for the case of the negative $\gamma$-characteristics $\gamma_{i}^{}<0$ ($i=1,\dots,5$) all $r$-matrices $r_{i}^{}$ $(i=2,\ldots,5)$ in (\ref{rmr27}) are reduced to the standard formula $r_{st}^{}:=r_{1}^{}$ and in the case of the positive $\gamma$-characteristics $\gamma_{i}^{}>0$ ($i=6,7$) the classical $r$-matrix $r_{7}^{}$ in (\ref{rmr28}) is reduced to the quasi-standard $r$-matrix $r_{qst}^{}:=r_{6}^{}$. 

Let us show that the $r$-matrix $r_{qst}^{}$ effectively depend only on one positive parameter. Indeed, it is easy to see that
\begin{eqnarray}
\begin{array}{rcl}\label{rmr29}
&&r_{qst}^{}=\beta_{+}'''r_{+}^{}\!+\beta_{-}'''r_{-}^{}=\imath\alpha(\lambda r_{+}^{}\!+\lambda^{-1}r_{-}^{})
\\[3pt]
&&\phantom{r_{qst}^{}}=\imath\alpha\big((\varphi\otimes\varphi)r_{+}^{}+(\varphi\otimes\varphi)r_{-}^{}\big), 
\end{array}
\end{eqnarray}
where $\varphi$ is the $\mathfrak{osp}^{\dag}(1|\mathfrak{sl}(2,\mathbb{R}))$-real automorphism: $\varphi(E_{\pm})=\lambda^{\pm1}E_{\pm}$, $\varphi(H)=H$, $\varphi(v_{\pm})=\lambda^{\pm\frac{1}{2}}v_{\pm}$, and $\alpha:=\sqrt{\beta_{+}'''\beta_{-}'''}\in\mathbb{R}^{+}$, $\lambda:=-\imath\sqrt{\beta_{+}'''{/}\beta_{-}'''}\in\mathbb{R}$.

Finally we obtain the following result:\\
\textit{For the non-compact real form $\mathfrak{osp}^{\dag}(1|\mathfrak{sl}(2;\mathbb{R}))$ with graded ($\varepsilon=1$) or ungraded ($\varepsilon=\imath$) conjugation $(^{\dag})$, there exists up to $\mathfrak{osp}^{\dag}(1|\mathfrak{sl}(2;\mathbb{R}))$-automorphisms four solutions of CYBE, namely Jordanian $r_{J}$, super-Jordanian $r_{sJ}$, standard $r_{st}$ and quasi-standard $r_{qst}$:
\begin{eqnarray}
&&\begin{array}{rcl}\label{rmr30}
&&r_{J}=\imath\beta E_{+}\wedge H,\;\;[r_{J},r_{J}]_{S}=0,
\end{array}
\\[5pt]
&&\begin{array}{rcl}\label{rmr31}
&&r_{sJ}=\imath\beta(E_{+}\wedge H+v_{+}\wedge v_{+}),
\\[2pt]
&&[r_{sJ},r_{sJ}]_{S}=0,
\end{array}
\\[5pt]
&&\begin{array}{rcl}\label{rmr32}
&&r_{st}=\imath\alpha(E_{+}\wedge E_{-}+2v_{+}\wedge v_{-}),
\\[2pt]
&&[r_{st}^{},r_{st}^{}]_{S}=-\alpha^{2}\Omega,
\end{array}
\end{eqnarray}
\\[-1cm]
\begin{eqnarray}
&&\begin{array}{rcl}\label{rmr33}
&&r_{qst}^{}=\imath\alpha(E_{+}\wedge H+v_{+}\wedge v_{+}
\\[2pt]
&&\phantom{r_{qst}^{}=}-E_{-}\wedge H-v_{-}\wedge v_{-}),
\\[2pt]
&&[r_{qst}^{},r_{qst}^{}]_{S}=\alpha^{2}\Omega,
\end{array}
\end{eqnarray}
where $\beta$ and $\alpha$ are positive numbers.}

III. {\large\textit{The non-compact real form}} $\mathfrak{osp}^{\dag}(1|\mathfrak{su}(1,1))$ (${H'}^{\dag}=H'$, ${E_{\pm}'}^{\dag}=-E_{\mp}'$, ${v_{\pm}'}^{\dag}=-\imath\varepsilon v_{\mp}'$).\\ 
It is not difficult to see that in the case of the grade ($\varepsilon=1$) and ungrade ($\varepsilon=\imath$) conjugations the basic two-tensors (\ref{rm5}) and (\ref{rm6}) have the following reality properties
\begin{eqnarray}\label{rmr34}
&&(r_{0}')^{\dag\otimes\dag}=-r_{0}',\quad (r_{\pm}')^{\dag\otimes\dag}=r_{\mp}',
\\[3pt]\label{rmr35}
&&(\bar{r}_{0}')^{\dag\otimes\dag}=-\bar{r}_{0}',\quad (\bar{r}_{\pm}')^{\dag\otimes\dag}=\bar{r}_{\mp}',
\end{eqnarray}
where the primed basic two-tensors $r_{i}'$ and $\bar{r}_{i}'$, ($i\in\{+,-,\,0\}$), are given  the formulas (\ref{rm5}) and (\ref{rm6}) in which the generators $E_{\pm}$, $H$, $v_{\pm}$ are replaced by the primed generators $E_{\pm}'$, $H'$, $v_{\pm}'$.
In the given case it follows from (\ref{rmr10}) that
\begin{eqnarray}\label{rmr36}
\beta_{0}^{*}=\beta_{0}^{},\quad\beta_{\pm}^{*}=-\beta_{\mp}^{}.
\end{eqnarray}

If $\beta_{0}^{2}+\beta_{+}^{}\beta_{-}^{}=0$ in (\ref{rm7}) and (\ref{rm9}) then $\beta_{0}^{}\beta_{0}^{*}-\beta_{\pm}^{}\beta_{\pm}^{*}=0$, i.e. $\beta_{\pm}^{}=\pm e^{\pm\imath\phi}|\beta_{0}^{}|$, and we have the following two $\phi$-families of $\mathfrak{osp}^{*}(1|\mathfrak{su}(1,1))$ homogeneous CYBE solutions:
\begin{eqnarray}\label{rmr37}
&&r_{\phi}':=\beta_{0}^{}\Bigl(e^{\imath\phi}\frac{|\beta_{0}^{}|}{\beta_{0}^{}}r_{+}'+r_{0}'-e^{-\imath\phi}\frac{|\beta_{0}^{}|}{\beta_{0}^{}}r_{-}'\Bigr),
\\[1pt]\label{rmr38}
&&\bar{r}_{\phi}':=\beta_{0}^{}\Bigl(e^{\imath\phi}\frac{|\beta_{0}^{}|}{\beta_{0}^{}}\bar{r}_{+}'+\bar{r}_{0}'-e^{-\imath\phi}\frac{|\beta_{0}^{}|}{\beta_{0}^{}}\bar{r}_{-}'\Bigr),
\end{eqnarray}
where $\beta_{0}^{}$ is real. By using the $\mathfrak{osp}^{*}(1|\mathfrak{su}(1,1))$-real rescaling automorphism $\varphi(E_{\pm}')=\Bigl(-\imath e^{\imath\phi}\frac{|\beta_{0}^{}|}{\beta_{0}^{}}\Bigr)^{\pm1}E_{\pm}'$, $\varphi(H')=H'$, $\varphi(v_{\pm}')=\Bigl(-\imath e^{\imath\phi}\frac{|\beta_{0}^{}|}{\beta_{0}^{}}\Bigr)^{\pm\frac{1}{2}}v_{\pm}$ we can reduce the $\phi$-families (\ref{rmr37}) and (\ref{rmr38}) to $r_{qsJ}':=\beta_{0}^{}(\imath r_{+}'+r_{0}'+\imath r_{-}')$ and $r_{qJ}':=\beta_{0}^{}(\imath\bar{r}_{+}'+\bar{r}_{0}'+\imath\bar{r}_{-}')$, respectively. Namely, we have
\begin{eqnarray}\label{rmr39}
\begin{array}{rcl}
&&r_{\phi}'=\displaystyle\beta_{0}^{}\Bigl(e^{\imath\phi}\frac{|\beta_{0}^{}|}{\beta_{0}^{}}r_{+}'+r_{0}'-e^{-\imath\phi}\frac{|\beta_{0}^{}|}{\beta_{0}^{}}r_{-}'\Big)
\\[10pt]
&&\phantom{r_{\phi}'}=\beta_{0}^{}\bigl(\imath(\varphi\otimes\varphi)r_{+}'+(\varphi\otimes\varphi)r_{0}'+\imath(\varphi\otimes\varphi)r_{-}')\bigr).
\end{array}
\end{eqnarray}
The same formula is also valid for the classical $r$-matrix $\bar{r}_{\phi}'$, (\ref{rmr38}). We shall call a $\mathfrak{osp}^{*}(1|\mathfrak{su}(1,1))$-antireal $r$-matrix of the form (\ref{rmr37}) (or (\ref{rmr38})) as ''quasi-super-Jordanian'' (or ''quasi-Jordanian'') if it can not be reduced to super-Jordanian (or Jordanian) form by a $\mathfrak{su}(1,1)$-real automorphism, but after complexification of $\mathfrak{osp}^{*}(1|\mathfrak{su}(1,1))$ it can be reduced to super-Jordanian (or Jordanian) form by an appropriate complex $\mathfrak{osp}(1|\mathfrak{sl}(2,\mathbb{C})$-automorphism. Thus all $r$-matrices in the $\phi$-families (\ref{rmr37})  (or (\ref{rmr38})) are quasi-super-Jordanian (or quasi-Jordanian)  and they are connected with each other by the $\mathfrak{osp}^{*}(1|\mathfrak{su}(1,1))$-real rescaling automorphism. We take $r_{qsJ}'$ and $r_{qJ}'$ as representatives of the $\phi$-families (\ref{rmr37}) and (\ref{rmr38}). It is easy to see that the $r$-matrices $r_{qsJ}'$ and $r_{qJ}'$ effectively depend only on positive values of the parameter $\beta_{0}^{}$. Indeed, we have
\begin{eqnarray}\label{rmr40}
\begin{array}{rcl}
&&r_{qsJ}'=\beta_{0}^{}(\imath r_{+}'+r_{0}'+\imath r_{-}')
\\[5pt]
&&\;\;=-\beta_{0}^{}\bigl(\imath(\varphi\otimes\varphi)r_{+}'\!+(\varphi\otimes\varphi)r_{0}'\!+\imath(\varphi\otimes\varphi)r_{-}'\bigr),
\end{array}
\end{eqnarray}
where $\varphi$ is the simple $\mathfrak{osp}^{*}(1|\mathfrak{su}(1,1))$ automorphism $\varphi(E_{\pm})=E_{\mp}$, $\varphi(H)=-H$, $\varphi(v_{\pm})=v_{\mp}$, i.e. any negative value of parameter $\beta_{0}^{}$ in $r_{qsJ}^{}$ can be replaced by a positive one. The same result is also valid for $r_{qJ}^{}$.
 
In the case $\beta_{0}^{2}+\beta_{+}^{}\beta_{-}^{}\neq0$ in (\ref{rm7}) we have four versions of $\mathfrak{osp}^{*}(1|\mathfrak{su}(1,1))$-antireal classical $r$-matrices. Two of them are characterized by positive value of $\gamma_{i}$, ($i=1,2$):
\begin{eqnarray}\label{rmr41}
\begin{array}{rcl}
&&r_{1}'=\beta_{0}^{}r_{0}',
\\[4pt]
&&r_{2}'=\beta_{+}'r_{+}'+\beta_{0}'r_{0}'-{\beta_{+}'}^{\!\!*}r_{-}',
\\[4pt]
&&[r_{i}',r_{i}']_{S}=\gamma_{i}\Omega\quad (i=1,2),
\end{array}
\end{eqnarray}
where $\beta_{0}^{}$ and $\beta_{0}'$ are real (see (\ref{rmr33})), and $\gamma_{1}=\beta_{0}^{2}>0$, $\gamma_{2}=\beta_{0}'{\beta_{0}'}^{*}-\beta_{+}'{\beta_{+}'}^{*}>0$. The remaining two are with negative values of $\gamma_{i}$, ($i=3,4$):
\begin{eqnarray}\label{rmr42}
\begin{array}{rcl}
&&r_{3}'=\beta_{+}''r_{+}'-{\beta_{+}''}^{\!*}r_{-}',
\\[4pt]
&&r_{4}'=\beta_{+}'''r_{+}'+\beta_{0}'''r_{0}'-{\beta_{+}'''}^{*}r_{-}',
\\[4pt]
&&[r_{i}',r_{i}']_{S}=\gamma_{i}\Omega\quad (i=3,4),
\end{array}
\end{eqnarray}
where $\beta_{0}'''$ is real (see (\ref{rmr33})), and $\gamma_{3}=-\beta_{+}''{\beta_{+}''}^{*}<0$, $\gamma_{4}=\beta_{0}'''{\beta_{0}'''}^{*}-\beta_{+}'''{\beta_{+}'''}^{*}<0$.

Let the classical $r$-matrices (\ref{rm26}) be $\mathfrak{su}(1,1)$-antireal, i.e. their parameters satisfy the reality conditions (\ref{rmr36}). In such case the functions (\ref{rm28}) for $\chi=e^{\imath\phi}$ have the same conjugation properties, i.e. $\tilde{\beta}_{0}^{*}=\tilde{\beta}_{0}^{}$, $\tilde{\beta}_{\pm}^{*}=-\tilde{\beta}_{\mp}^{}$, and we obtain that the automorphism (\ref{rm25}) with these parameters is $\mathfrak{su}(1,1)$-real, i.e.  
\begin{eqnarray}\label{rmr43}
\begin{array}{rcl}
&&(\varphi_{1}^{}(E_{\pm}'))^{\dag}=\varphi_{1}^{}((E_{\pm}')^{\dag})=-\varphi_{1}^{}(E_{\mp}'),
\\[4pt] 
&&(\varphi_{1}^{}(H'))^{\dag}=\varphi_{1}^{}((H')^{\dag})=\varphi_{1}^{}(H'),         
\\[4pt]
&&(\varphi_{1}^{}(v_{\pm}')^{\dag}=\varphi_{1}^{}((v_{\pm}')^{\dag})=-\imath\varepsilon\varphi_{1}^{}(v_{\mp}').
\end{array}
\end{eqnarray}
It allows to reduce the $r$-matrix $r_{2}'$ to the standard $r$-matrix $r_{st}':=r_{1}'$ for $\gamma_{1}^{}=\gamma_{2}^{}>0$ and the $r$-matrix $r_{3}'$ to the $r$-matrix $r_{4}'$  for $\gamma_{3}^{}=\gamma_{4}^{}<0$ by use of the formula (\ref{rm16}). By analogy to the notation of quasi-Jordanian $r$-matrix we shall call the $r$-matrices $r_{3}'$ and $r_{4}'$ as quasi-standard ones and take $r_{qst}'=\alpha(E_{+}'\wedge H'+v_{+}'\wedge v_{+}'+E_{-}'\wedge H'+v_{-}'\wedge v_{-}')$ as their representative\footnote{The $r$-matrix $r_{qst}'$ is connected with $r_{3}'$ (\ref{rmr42}) by the following way. Substituting $\beta_{+}^{}=|\beta_{+}^{}|e^{\imath\phi}$ in $r_{3}'$ (\ref{rmr42}) and using the $\mathfrak{su}(1,1)$-real rescaling automorphism $\varphi(E_{\pm}')=e^{\pm\imath\phi}E_{\pm}'$, $\varphi(H')=H'$, $\varphi(v_{\pm}')=e^{\pm\frac{\imath\phi}{2}}v_{\pm}'$ we obtain $r_{qst}'$ with $\alpha=|\beta_{+}^{}|$.}.

Finally for $\mathfrak{su}(1,1)$ we obtain:\\
\textit{For the non-compact real form $\mathfrak{osp}^{\dag}(1|\mathfrak{su}(1,1))$ with graded ($\varepsilon=1$) or ungraded ($\varepsilon=\imath$) conjugation $(^{\dag})$, there exists up to $\mathfrak{osp}^{\dag}(1|\mathfrak{su}(1,1))$-automorphisms four solutions of CYBE, namely quasi-Jordanian $r_{qJ}'$, quasi-super-Jordanian $r_{qsJ}'$, quasi-standard $r_{qst}'$ and standard $r_{st}'$:
\begin{eqnarray}
&&\begin{array}{rcl}\label{rmr44}
&&r_{qJ}'=\beta\bigl(\imath(E_{+}'-E_{-}')\wedge H'+E_{+}'\wedge E_{-}'\bigr),
\\[4pt]
&&[r_{qJ}',r_{qJ}']_{S}=0,
\end{array}
\\[7pt]
&&\begin{array}{rcl}\label{rmr45}
&&r_{qsJ}'=\beta\bigl(\imath((E_{+}'-E_{-}')\wedge H'\!+v_{+}'\wedge v_{+}'-v_{-}'\wedge v_{-}')
\\[3pt]
&&\phantom{r_{qsJ}'=}+E_{+}'\wedge E_{-}'\!+2v_{+}'\wedge v_{-}'\bigl),
\\[4pt]
&&[r_{qsJ}',r_{qsJ}']_{S}=0,
\end{array}
\\[7pt]
&&\begin{array}{rcl}\label{rmr46}
&&r_{qst}'=\alpha(E_{+}'\wedge H'\!+v_{+}'\wedge v_{+}'
\\[3pt]
&&\phantom{r_{qst}'=}+E_{-}'\wedge H'+v_{-}'\wedge v_{-}'),
\\[4pt]
&&[r_{qst}',r_{qst}']_{S}=-\alpha^{2}\Omega,
\end{array}
\end{eqnarray}
\\[-1cm]
\begin{eqnarray}
&&\begin{array}{rcl}\label{rmr47}
&&r_{st}'=\alpha(E_{+}'\wedge E_{-}'+2v_{+}'\wedge v_{-}'),
\\[4pt]
&&[r_{st}',r_{st}']_{S}=\alpha^{2}\Omega,
\end{array}
\end{eqnarray}
where $\beta$ and $\alpha$ are positive numbers}.

\setcounter{equation}{0}
\section{Isomorphism between $\mathfrak{osp}^{\dag}(1|\mathfrak{sl}(2;\mathbb{R}))$ and $\mathfrak{osp}^{\dag}(1|\mathfrak{su}(1,1))$ \\ bisuperalgebras and its application to quantizations \\ of $N=1$, $D=3$ Lorentz supersymmetry $\mathfrak{osp}^{\dag}(1|\mathfrak{o}(2,1))$} 
Using the formulas of connections between the CW and Cartesian bases (see (\ref{pr11'}), (\ref{pr13})) we can express the classical $\mathfrak{osp}^{\dag}(1|\mathfrak{sl}(2;\mathbb{R}))$ and $\mathfrak{osp}^{\dag}(1|\mathfrak{su}(1,1))$ $r$-matrices in terms of the $\mathfrak{osp}^{\dag}(1|\mathfrak{o}(2,1))$ Cartesian basis. We get the following results. 

For the non-compact real form $\mathfrak{osp}^{\dag}(1|\mathfrak{sl}(2;\mathbb{R}))$ with graded and ungraded conjugation $(^{\dag})$:
\begin{eqnarray}\label{is1}
&&\begin{array}{rcl}
&&r_{J}=\imath\beta E_{+}\wedge H
\\[3pt]
&&\phantom{r_{J}}=-\beta(\imath I_{1}-I_{2})\wedge I_{3},
\\[4pt]
&&[r_{J},r_{J}]_{S}\;=\;0,
\end{array}
\\[7pt]
&&\begin{array}{rcl}\label{is2}
&&r_{sJ}=\imath\beta(E_{+}\wedge H+v_{+}\wedge v_{+})
\\[3pt]
&&\phantom{r_{sJ}}=-\beta(\imath I_{1}-I_{2})\wedge I_{3}+v_{1}\wedge v_{1})
\\[4pt]
&&[r_{sJ},r_{sJ}]_{S}=0,
\end{array}
\end{eqnarray}
\begin{eqnarray}\label{is1}
&&\begin{array}{rcl}\label{is3}
&&r_{st}=\imath\alpha(E_{+}\wedge E_{-}+2v_{+}\wedge v_{-})
\\[3pt]
&&\phantom{r_{st}}=-2\alpha(I_{1}\wedge I_{2}+2v_{1}\wedge v_{2}),
\\[4pt]
&&[r_{st},r_{st}]_{S}=-\alpha^{2}\Omega,
\end{array}
\\[7pt]
&&\begin{array}{rcl}\label{is4}
&&r_{qst}=\imath\alpha(E_{+}\wedge H+v_{+}\wedge v_{+}
\\[3pt]
&&\phantom{r_{qst}=}+E_{-}\wedge H+v_{-}\wedge v_{-})
\\[3pt]
&&\phantom{r_{qst}}=-2\imath\alpha (I_{1}\wedge I_{3}+v_{1}\wedge v_{1}-v_{2}\wedge v_{2}),
\\[4pt]
&&[r_{qst},r_{qst}]_{S}=\alpha^{2}\Omega,
\end{array}
\end{eqnarray}
where $\beta$ and $\alpha$ are positive numbers. 

For the noncompact real form $\mathfrak{osp}^{\dag}(1|\mathfrak{su}(1,1))$ with graded and ungraded conjugation $(^{\dag})$:
\begin{eqnarray}
&&\begin{array}{rcl}\label{is5}
&&r_{qJ}'=\beta\bigl(\imath(E_{+}'-E_{-}')\wedge H'+E_{+}'\wedge E_{-}'\bigr),
\\[3pt]
&&\phantom{r_{qJ}'}=-\beta(\imath I_{1}-I_{2})\wedge I_{3},
\\[3pt]
&&[r_{qJ}',r_{qJ}']_{S}=0,
\end{array}
\\[7pt]
&&\begin{array}{rcl}\label{is6}
&&r_{qsJ}'=\beta\bigl(\imath((E_{+}'-E_{-}')\wedge H'\!+v_{+}'\wedge v_{+}'-v_{-}'\wedge v_{-}')
\\[3pt]
&&\phantom{r_{qsJ}'=}+E_{+}'\wedge E_{-}'\!+2v_{+}'\wedge v_{-}'\bigl),
\\[3pt]
&&\phantom{r_{qsJ}'}=-\beta(\imath I_{1}-I_{2})\wedge I_{3}+v_{1}\wedge v_{1}),
\\[3pt]
&&[r_{qsJ}',r_{qsJ}']_{S}=0,
\end{array}
\end{eqnarray}
\\[-1cm]
\begin{eqnarray}
&&\begin{array}{rcl}\label{is7}
&&r_{qst}'=\alpha(E_{+}'\wedge H'+v_{+}'\wedge v_{+}'
\\[3pt]
&&\phantom{r_{qst}'=}+E_{-}'\wedge H'+v_{-}'\wedge v_{-}')
\\[3pt]
&&\phantom{r_{qst}'}=-2\alpha(I_{1}\wedge I_{2}+2v_{1}\wedge v_{2}),
\\[4pt]
&&[r_{qst}',r_{qst}']_{S}=-\alpha^{2}\Omega,
\end{array}
\\[7pt]
&&\begin{array}{rcl}\label{is8}
&&r_{st}'=\alpha(E_{+}'\wedge E_{-}'+2v_{+}'\wedge v_{-}')
\\[4pt]
&&\phantom{r_{st}'}=-2\imath\alpha (I_{1}\wedge I_{3}+v_{1}\wedge v_{1}+zv_{2}\wedge v_{2}),
\\[4pt]
&&[r_{st}',r_{st}']_{S}=\alpha^{2}\Omega,
\end{array}
\end{eqnarray}
where $\beta$ and $\alpha$ are positive numbers.

Comparing the $r$-matrix expressions (\ref{is1})--(\ref{is4}) with (\ref{is5})-(\ref{is8}) we obtain that
\begin{eqnarray}\label{is9}
&&r_{J}=r_{qJ}'=-\alpha\big((\imath I_{1}-I_{2})\wedge I_{3}\big),
\\[3pt] \label{is10}
&&r_{sJ}=r_{qsJ}'=-\alpha\big((\imath I_{1}-I_{2})\wedge I_{3}+v_{1}\wedge v_{1}\big),
\\[3pt] \label{is11}
&&r_{st}=r_{qst}'=-2\alpha(I_{1}\wedge I_{2}+2v_{1}\wedge v_{2}),
\\[3pt] \label{is12}
&&r_{qst}=r_{st}'=-2\imath\alpha (I_{1}\wedge I_{3}+v_{1}\wedge v_{1}-v_{2}\wedge v_{2}).
\end{eqnarray}
We see the following:
\begin{itemize}
\item[(i)]
The Jordanian $r$-matrix $r_{J}$ in the $\mathfrak{osp}^{\dag}(1|\mathfrak{sl}(2;\mathbb{R}))$ basis is the same as the quasi-Jordanian $r$-matrix $r_{qJ}'$ in the $\mathfrak{osp}^{\dag}(1|\mathfrak{su}(1,1))$ basis. 
\item[(ii)]
The super-Jordanian $r$-matrix  $r_{sJ}$ in the $\mathfrak{osp}^{\dag}(1|\mathfrak{sl}(2;\mathbb{R}))$ basis is the same as the super-quasi-Jordanian $r$-matrix $r_{qsJ}'$ in the $\mathfrak{osp}^{\dag}(1|\mathfrak{su}(1,1))$ basis. 
\item[(iii)]
The standard $r$-matrix $r_{st}$ in the $\mathfrak{osp}^{\dag}(1|\mathfrak{sl}(2;\mathbb{R}))$ basis becomes the quasi-standard $r$-matrix $r_{qst}'$ in the $\mathfrak{osp}^{\dag}(1|\mathfrak{su}(1.1))$ basis. 
\item[(iv)]
Conversely, the quasi-standard $r$-matrix $r_{qst}$ in the $\mathfrak{osp}^{\dag}(1|\mathfrak{su}(2,\mathbb{R}))$ basis is the same as the standard $r$-matrix $r_{st}'$ in the $\mathfrak{osp}^{\dag}(1|\mathfrak{su}(1,1))$ basis.
\end{itemize}
\textit{The relations (\ref{is9})--(\ref{is12}) show that the $\mathfrak{osp}^{\dag}(1|\mathfrak{su}(2,\mathbb{R}))$ and $\mathfrak{osp}^{\dag}(1|\mathfrak{su}(1,1))$ superbialgebras are isomorphic}. This result finally resolves the doubts about isomorphisms of these two superbialgebras \cite{BoHaReSe2011})

Using the isomorphisms of the $\mathfrak{osp}^{\dag}(1|\mathfrak{su}(2,\mathbb{R}))$ and $\mathfrak{osp}^{\dag}(1|\mathfrak{su}(1,1))$ bialgebras we take as basic $r$-matrices for the $N=1$, $D=3$ Lorentz superalgebra $\mathfrak{osp}^{\dag}(1|\mathfrak{o}(2,1))$ the following ones:                 
\begin{eqnarray}
&&\begin{array}{rcl}\label{is13}
&&r_{J}=-\beta\big((\imath I_{1}-I_{2})\wedge I_{3}\big)
\\[3pt]
&&\phantom{r_{J}}=\imath\beta E_{+}\wedge H,
\end{array}
\\[7pt]
&&\begin{array}{rcl}\label{is14}
&&r_{sJ}=-\beta\big((\imath I_{1}-I_{2})\wedge I_{3}+v_{1}\wedge v_{1}\big)
\\[3pt] 
&&\phantom{r_{sJ}}=\imath\beta(E_{+}\wedge H+v_{+}\wedge v_{+}),
\end{array}
\\[7pt] 
&&\begin{array}{rcl}\label{is15}
&&r_{st}=-2\alpha(I_{1}\wedge I_{2}+2v_{1}\wedge v_{2})
\\[3pt] 
&&\phantom{r_{st}}=\imath\alpha(E_{+}\wedge E_{-}+2v_{+}\wedge v_{-}),
\end{array}
\\[7pt] 
&&\begin{array}{rcl}\label{is16}
&&r_{st}'=-2\imath\alpha(I_{1}\wedge I_{3}+v_{1}\wedge v_{1}-v_{2}\wedge v_{2})
\\[3pt] 
&&\phantom{r_{st}'}=\alpha(E_{+}'\wedge E_{-}'+2v_{+}'\wedge v_{-}'),
\end{array}
\end{eqnarray}
where $\beta$ and $\alpha$ are positive numbers, moreover the parameter $\alpha$ is  effective whereas the parameter $\beta$ is not effective, i.e. it can be removed by a $\mathfrak{osp}^{\dag}(1|\mathfrak{sl}(2;\mathbb{R}))$-real rescaling automorphism \footnote{Here we keep the non-effective parameter $\beta$ for convenient of quantization.}: $\varphi(E_{+})=\beta^{-1}E_{+}$, $\varphi(E_{-})=\beta E_{-}$, $\varphi(v_{+})=\sqrt{\beta^{-1}}\,v_{+}$, $\varphi(v_{-})=\sqrt{\beta}\,v_{-}$, $\varphi(H)=H$.

 The first two $r$-matrices $r_{J}$ and $r_{sJ}$ present the Jordanian and super-Jordanian twist deformations of $\mathfrak{osp}^{\dag}(1|\mathfrak{sl}(2;\mathbb{R}))$, the third and fourth $r$-matrices $r_{st}$ and $r_{st}'$ correspond to the $q$-analogs of $\mathfrak{osp}^{\dag}(1|\mathfrak{sl}(2;\mathbb{R}))$ and $\mathfrak{osp}^{\dag}(1|\mathfrak{su}(1,1))$ real algebras. In the next section we shall show how to quantize the $r$-matrices (\ref{is13})--(\ref{is16}) in an explicit form.

\setcounter{equation}{0}
\section{Quantizations of the $N=1$, $D=3$ Lorentz supersymmetry $\mathfrak{osp}^{\dag}(1|\mathfrak{o}(2,1))$}
Comparing the classical $r$-matrices of the complex Lie superalgebra $\mathfrak{osp}(1|2;\mathbb{C})$ (\ref{rm30})--(\ref{rm32}) with the classical $r$-matrices of its real forms: $\mathfrak{osp}^{*}(1|\mathfrak{o}(3))$ (\ref{rmr20}) and $\mathfrak{osp}^{\dag}(1|\mathfrak{o}(2,1))$ (\ref{is13})--(\ref{is16}), we see that they are given by the same formulas and differ in values of the deformation parameters and conjugation properties. There are the similar picture for the deformed structures of $\mathfrak{osp}(1|2;\mathbb{C})$ and its real forms. Therefore one can obtain the quantum deformations of all $\mathfrak{osp}(1|2;\mathbb{C})$ real forms from the quantum deformation of $\mathfrak{osp}(1|2;\mathbb{C})$ by specialization of the corresponding parameter deformation and the conjugation property. 

\textit{\large{1. $q$-Analogs}}. The quantum Hopf deformations corresponding to the standard classical $r$-matrices (\ref{rm32}), (\ref{rmr20}), (\ref{is15}) and (\ref{is16}) are called the $q$-analogs, The $q$-analog of $U(\mathfrak{g})$ ($\mathfrak{g}=\mathfrak{osp}(1|2;\mathbb{C}),\,\mathfrak{osp}^{*}(1|\mathfrak{su}(2)),\,\mathfrak{osp}^{\dag}(1|\mathfrak{sl}(2;\mathbb{R})),\,\mathfrak{osp}^{\dag}(1|\mathfrak{su}(1,1))$) is an unital associative algebra $U_{q}(\mathfrak{g})$ with the generators\footnote{They are $q$-analogs of the Chevalley basis $v_{\pm}$, $H$ with the defining relations: $\{v_{+},v_{-}\}=-\frac{1}{2}H$, $[H,v_{\pm}]=\pm\frac{1}{2}v_{\pm}$ (see (\ref{pr2})--(\ref{pr3})).} $x_{\pm}$, $q^{\pm\frac{1}{2}X_{0}}$ and the defining relations:
\begin{eqnarray}\label{qu1}
\begin{array}{rcl}
&&q^{\frac{1}{2}X_{0}}q^{-\frac{1}{2}X_{0}}=q^{-\frac{1}{2}X_{0}}q^{\frac{1}{2}X_{0}}=1,
\\[4pt] 
&&q^{\frac{1}{2}X_{0}}x_{\pm}=q^{\pm\frac{1}{4}}x_{\pm}q^{\frac{1}{2}X_{0}},
\\[4pt] 
&&\{x_{+},x_{-}\}=\displaystyle\frac{q^{-\frac{1}{2}X_{0}}-q^{\frac{1}{2}X_{0}}}{q-q^{-1}},
\end{array}
\end{eqnarray} with the 
additional conditions:
\begin{eqnarray}\label{qu2}
\begin{array}{rcl}
&&(a)\;q=e^{\beta}\;(\beta\in\mathbb{C})\;{\rm for}\;U_{q}(\mathfrak{osp}^{}(1|2;\mathbb{C})),
\\[7pt]
&&(b)\;x_{\pm}^{*}=\pm\varepsilon x_{\mp}^{},\;(q^{\frac{1}{2}X_0})^{*}=q^{\frac{1}{2}X_0},\;q=e^{\alpha}
\\[2pt]
&&\phantom{aaa}{\rm for}\;U_{q}(\mathfrak{osp}^{*}(1|\mathfrak{su}(2))),
\\[7pt]
&&(c)\;x_{\pm}^{\dag}=-\varepsilon x_{\pm}^{},\;(q^{\frac{1}{2}X_0})^{\dag}=q^{\frac{1}{2}X_0},\;q=e^{\imath\alpha}
\\[2pt]
&&\phantom{aaa}{\rm for}\;U_{q}(\mathfrak{osp}^{\dag}(1|\mathfrak{sl}(2;\mathbb{R})))\simeq U_{q}(\mathfrak{osp}^{\dag}(1|\mathfrak{o}(2,1))),
\\[7pt]
&&(c')\;x_{\pm}^{\dag}=\imath\varepsilon x_{\mp}^{},\; (q^{\frac{1}{2}X_0})^{\dag}=q^{\frac{1}{2}X_0},\;\; q=e^{\alpha}
\\[2pt]
&&\phantom{aaa}{\rm for}\;U_{q}(\mathfrak{osp}^{\dag}(1|\mathfrak{su}(1,1)))\simeq U_{q}(\mathfrak{osp}^{\dag}(1|\mathfrak{o}(2,1))),
\end{array}
\end{eqnarray}
where $\alpha$ is real in accordance with (\ref{rmr20}), (\ref{is15}) and (\ref{is16}), for the graded ($\varepsilon=1$) or ungraded ($\varepsilon=\imath$) conjugation $(^{\divideontimes})$, ($\divideontimes=*,\dag$) (see (\ref{rmr6}), (\ref{rmr7}).  

A Hopf structure on $U_{q}(\mathfrak{g})$ is defined with help of three additional operations: coproduct (comultiplication) $\Delta_{q}$, antipode $S_{q}$ and counit $\epsilon_{q}$:
\begin{eqnarray}\label{qu3}
\begin{array}{rcl}
&&\Delta_{q}(q^{\pm\frac{1}{2}X_{0}})=q^{\pm\frac{1}{2}X_{0}}\otimes q^{\pm\frac{1}{2}X_{0}},
\\[4pt] 
&&\Delta_{q}(x_{\pm}^{})=x_{\pm}^{}\otimes q^{\frac{1}{4}X_{0}}+q^{-\frac{1}{4}X_{0}}\otimes x_{\pm}^{},
\\[4pt]
&&S_{q}(q^{\pm\frac{1}{2}X_{0}})=q^{\mp\frac{1}{2}X_{0}},\;\; S_{q}(x_{\pm}^{})=-q^{\pm\frac{1}{4}}x_{\pm}^{},
\\[4pt]
&&\epsilon_{q}(q^{\pm\frac{1}{2}X_{0}})=1,\;\;\epsilon_{q}(x_{\pm}^{})=0,
\end{array}
\end{eqnarray}
with the reality conditions for the real form $U_q(\mathfrak{g}^{\divideontimes})$ ($\divideontimes=*,\dag$):
\begin{eqnarray}\label{qu4}
\begin{array}{rcl}
&&(\Delta_{q}(X))^{\divideontimes\otimes\divideontimes}=\Delta_{q}(X^{\divideontimes}),
\\[4pt]
&&(S_{q}(X))^{\divideontimes}=S_{q}^{-1}(X^{\divideontimes}),\;\;(\epsilon_{q}(X))^{\divideontimes}=\epsilon_{q}(X^{\divideontimes})
\end{array}
\end{eqnarray}
for any $X\in U_{q}(\mathfrak{g}^{\divideontimes})$ ($\mathfrak{g}^{\divideontimes}=\mathfrak{osp}^{*}(1|\mathfrak{su}(2)),\,\mathfrak{osp}^{\dag}(1|\mathfrak{sl}(2;\mathbb{R})),\,\mathfrak{osp}^{\dag}(1|\mathfrak{su}(1,1))$). 

The quantum algebra $U_{q}(\mathfrak{g})$ is endowed also with the {\it opposite} Hopf structure: opposite  coproduct $\tilde{\Delta}_{q}$\footnote{The opposite coproduct $\tilde{\Delta}_{q}(\cdot)$ is a coproduct with permuted components, i.e. $\tilde{\Delta}_{q}(\cdot)=\tau\circ\Delta_{q}(\cdot)$ where $\tau$ is the super flip operator (see \ref{rmr4}).}, corresponding antipode $\tilde{S}_{q}$ and counit $\tilde{\epsilon}_{q}$.

An invertible element $R_{q}:=R_{q}(\mathfrak{g})$ which satisfies the relations: 
\begin{eqnarray}\label{qu5}
\begin{array}{rcl}
&&R_{q}\Delta_{q}(X)=\tilde{\Delta}_{q}(X)R_{q},\;\forall X\in U_{q}(\mathfrak{g}),  
\\[5pt]
&&(\Delta_{q}\otimes{\rm id})R_{q}=R_{q}^{13}R_{q}^{23},
\\[5pt]
&&({\rm id}\otimes\Delta_{q})R_{q}=R_{q}^{12}R_{q}^{13}
\end{array}
\end{eqnarray}
as well as, due to (\ref{qu5}), the quantum Yang-Baxter equation (QYBE)
\begin{eqnarray}\label{qu6}
\begin{array}{rcl}
&&R_{q}^{12}R_{q}^{13}R_{q}^{23}=R_{q}^{23}R_{q}^{13}R_{q}^{12}
\end{array}
\end{eqnarray}
is called the {\it universal $R$-matrix}. 

Let $U_{q}(\mathfrak{b}_{+})$ and  $U_{q}(\mathfrak{b}_{-})$ be quantum Borel subalgebras of $U_{q}(\mathfrak{g})$, generated by $x_{+}$, $q^{\pm\frac{1}{4}X_{0}}$ and $x_{-}$, $q^{\pm\frac{1}{4}X_{0}}$ respectively. We denote by $T_{q}(\mathfrak{b}_{+}\otimes\mathfrak{b}_{-})$ the Taylor extension of $U_{q}(\mathfrak{b}_{+})\otimes U_{q}(\mathfrak{b}_{-})$\footnote{$T_{q}(\mathfrak{b}_{+}\otimes\mathfrak{b}_{-})$  is an associative algebra generated by formal Taylor series of the monomials $x_{+}^{n}\otimes x_{-}^{m}$ with coefficients which are rational functions of $q^{\pm\frac{1}{4}X_{0}}$, $q^{\pm X_{0}\otimes X_{0}}$, provided that all values $|n-m|$ for each formal series are bounded, $|n-m|<N$.}. One can show (see \cite{KhTo1991,KhTo1992}) that \textit{there exists unique solution of equations (\ref{qu5}) in the space $T_{q}(\mathfrak{b}_{+}\otimes\mathfrak{b}_{-})$ and such solution has the following form}\footnote{The formulas (\ref{qu7})--(\ref{qu9}) are a specialization of the formulas (7.1)-(7.3) from the article \cite{KhTo1991} to the case (\ref{qu1}).}
\begin{eqnarray}\label{qu7}
\begin{array}{rcl}
&&R_{q}(\mathfrak{g}):=R_{q}^{\succ}=\check{R}_{q}^{\succ}K,
\end{array}
\end{eqnarray}
where
\begin{eqnarray}\label{qu8}
&&\begin{array}{rcl}
K=q^{X_{0}\otimes X_{0}},
\end{array}
\\[4pt]
&&\begin{array}{rcl}\label{qu9}
\check{R}_{q}^{\succ}=\exp_{\tilde{q}}\big((q^{-1}-q)x_{+}q^{-\frac{1}{4}X_{0}}\otimes q^{\frac{1}{4}X_{0}}x_{-}\big), 
\end{array}
\end{eqnarray}
Here $\tilde{q}:=-q^{-\frac{1}{4}}$ and the deformation parameter $q$ is given by the conditions (\ref{qu2}). We also use the standard definition of the $\tilde{q}$-exponential: 
\begin{eqnarray}\label{qu10}
\begin{array}{rcl}
&&\exp_{\tilde{q}}(x):=\displaystyle\sum_{n\geq0}\frac{x^{n}}{(n)_{\tilde{q}}!},\;\;(n)_{\tilde{q}}:=\frac{(1-\tilde{q}^{n})}{(1-\tilde{q})}, 
\\[15pt]
&&(n)_{\tilde{q}}!:=(1)_{\tilde{q}}(2)_{\tilde{q}}\dots(n)_{\tilde{q}}.
\end{array} 
\end{eqnarray}
Analogously, \textit{there exits unique solution of equations (\ref{qu5}) in the space $T_{q}(\mathfrak{b}_{-}\otimes\mathfrak{b}_{+})=\tau\circ T_{q}(\mathfrak{b}_{+}\otimes\mathfrak{b}_{-})$ and such solution is given by the formula 
\begin{eqnarray}\label{qu11}
\begin{array}{rcl}
&&R_{q}(\mathfrak{g}):=R_{q}^{\prec}=\check{R}_{q}^{\prec}K^{-1},
\end{array}
\end{eqnarray} 
where $K$ is given by the formula (\ref{qu8}), and}
\begin{eqnarray}\label{qu12}
&&\begin{array}{rcl}
\check{R}_{q}^{\prec}=\exp_{\tilde{q}'}\big((q^{-1}-q)x_{-}q^{\frac{1}{4}X_{0}}\otimes q^{-\frac{1}{4}X_{0}}x_{+}\big). 
\end{array}
\end{eqnarray}
Here $\tilde{q}'=\tilde{q}^{-1}=-q^{\frac{1}{4}}$, and $q$ satisfies the conditions (\ref{qu2}).

As formal Taylor series the solutions (\ref{qu7})--(\ref{qu9}) and (\ref{qu11}), (\ref{qu12}) are independent and they are related by 
\begin{eqnarray}\label{qu13}
&&R_{q}^{\prec}=\tau\circ R_{q^{-1}}^{\succ}.
\end{eqnarray}
It should be noted also that 
\begin{eqnarray}\label{qu14}
&&(R_{q}^{\succ})^{-1}=R_{q^{-1}}^{\succ},\;\;(R_{q}^{\prec})^{-1}=R_{q^{-1}}^{\prec}
\end{eqnarray}
for all quantum superalgebras $U_{q}(\mathfrak{g})$, $\mathfrak{g}=\mathfrak{osp}(1|2;\mathbb{C})$, $\mathfrak{osp}^{*}(1|\mathfrak{su}(2))$, $\mathfrak{osp}^{\dag}(1|\mathfrak{sl}(2;\mathbb{R}))$, $\mathfrak{osp}^{\dag}(1|\mathfrak{su}(1,1))$. 
From the explicite forms (\ref{qu7})--(\ref{qu9}) and (\ref{qu11}), (\ref{qu12}) we also see that ($\divideontimes=*,\dag$):
\begin{eqnarray}\label{qu15}
\begin{array}{rcl}
&&(R_{q}^{\succ})^{\divideontimes}=\tau\circ R_{q}^{\succ}=(R_{q}^{\prec})^{{-1}},
\\[5pt]
&&(R_{q}^{\prec})^{\divideontimes}=\tau\circ R_{q}^{\prec}=(R_{q}^{\succ})^{-1}
\end{array}
\end{eqnarray}
for $U_{q}(\mathfrak{osp}^{*}(1|\mathfrak{su}(2)))\simeq U_{q}(\mathfrak{osp}^{*}(1|\mathfrak{0}(3)))$
and $U_{q}(\mathfrak{osp}^{\dag}(1|\mathfrak{su}(1,1)))\simeq U_{q}(\mathfrak{osp}^{\dag}(1|\mathfrak{o}(2,1)))$, and
\begin{eqnarray}\label{qu16}
\begin{array}{rcl}
&&(R_{q}^{\succ})^{\dag}=(R_{q}^{\succ})^{{-1}},\;\;(R_{q}^{\prec})^{\dag}=(R_{q}^{\prec})^{-1}
\end{array}
\end{eqnarray}
for $U_{q}(\mathfrak{osp}^{\dag}(1|\mathfrak{sl}(2;\mathbb{R})))\simeq U_{q}(\mathfrak{osp}^{\dag}(1|\mathfrak{o}(2,1)))$.
Thus, in the case $U_{q}(\mathfrak{osp}^{\dag}(1|\mathfrak{sl}(2;\mathbb{R})))$ both $R$-matrices $R_{q}^{\succ}$, $R_{q}^{\prec}$ are unitary and in the case $U_{q}(\mathfrak{osp}^{*}(1|\mathfrak{su}(2)))$ and $U_{q}(\mathfrak{osp}^{\dag}(1|\mathfrak{su}(1,1)))$ they can be called ''$\tau$-Hermitian''. 

In the limit $\alpha\rightarrow0$ ($q\rightarrow1$) we obtain for the $R$-matrix (\ref{qu7})--(\ref{qu9})
\begin{eqnarray}\label{qu17}
\begin{array}{rcl}
&&R_{q}(\mathfrak{g})=1+r_{BD}^{}+\textit{O}(\alpha^{2}).
\end{array}
\end{eqnarray}
Here $r_{BD}^{}$ is the classical Belavin-Drinfeld $r$-matrix:
\begin{eqnarray}\label{qu18}
\begin{array}{rcl}
&&r_{BD}^{}=\beta\bigl(2\tilde{x}_{+}^{}\otimes\tilde{x}_{-}^{}-16\tilde{x}_{+}^{2}\otimes \tilde{x}_{-}^{2}+X_{0}^{}\otimes X_{0}^{}\bigr),
\end{array}
\end{eqnarray}
where $\beta=\ln{q}$ (see (\ref{qu2})), and\footnote{In the case of $\mathfrak{osp}^{\dag}(1|\mathfrak{su}(1,1))$ we need to use the primed CW basis $\tilde{x}_{\pm}=E_{\pm}'$, $X_{0}=H'$ (see (\ref{pr13}), (\ref{pr14})).} $\tilde{x}_{\pm}^{}=v_{\pm}^{}$, $\tilde{x}_{\pm}^{2}=\pm\frac{1}{4}E_{\pm}^{}$, $X_{0}^{}=H$.
The $r$-matrix $r_{BD}^{}$ is not skew-symmetric and it satisfies the standard CYBE 
\begin{eqnarray}\label{qu19}
[r_{BD}^{12},r_{BD}^{13}+r_{BD}^{23}]+[r_{BD}^{13},r_{BD}^{23}]=0 
\end{eqnarray}
which is obtained from QYBE (\ref{qu6}) in the limit (\ref{qu17}). The standard $r$-matrix (\ref{is11}) or (\ref{is12}) is  the skew-symmetric part of $r_{BD}^{}$, namely
\begin{eqnarray}\label{qu20}
&&r_{BD}^{}=\frac{1}{2}\bar{r}_{st}^{}+\frac{1}{2}\bar{C}_{2},
\end{eqnarray}
where $\bar{r}_{st}^{}=r_{BD}^{}-\tau\circ r_{BD}^{}$ is the standard $r$-matrix (\ref{rm32}), (\ref{rmr20}),     (\ref{is15}) or (\ref{is16}), and $\bar{C}_{2}=2\beta C_{2}=r_{BD}^{}+\tau\circ r_{BD}^{}$ where $C_{2}$ is the split Casimir element of $\mathfrak{g}$, ($\mathfrak{g}=\mathfrak{osp}(1|2;\mathbb{C})$, $\mathfrak{osp}^{*}(1|\mathfrak{su}(2))$, $\mathfrak{osp}^{\dag}(1|\mathfrak{sl}(2;\mathbb{R}))$, $\mathfrak{osp}^{\dag}(1|\mathfrak{su}(1,1))$).

\textit{\large{2. Twisted deformations}}. The classical $r$-matrices (\ref{rm30}), (\ref{rm31}) and (\ref{is13}), (\ref{is14}) satisfy the homogeneous CYBE (\ref{rm1}) with the vanishing right side and the corresponding Hopf deformations are determined by twisting two-tensors. We remind basic properties of the twisted deformation of a Hopf (super)algebra \cite{Dr1985}. Let $\mathcal{A}:=\mathcal{A}(A;m,\Delta,S,\varepsilon$) be a Hopf (super)algebra with multiplication $m$, coproduct $\Delta$, antipode $S$ and counit $\epsilon$ and let $F\in{A}\otimes{A}$ be an invertible two-tensor satisfies the 2-cocycle condition
\begin{equation}\label{qu21}
F^{12}(\Delta\otimes{\rm id})(F)=F^{23}({\rm id}\otimes\Delta)(F)
\end{equation}
and the "unital" normalization
\begin{equation}\label{qu22}
(\epsilon\otimes{\rm id})(F)=({\rm id}\otimes\epsilon )(F)=1.
\end{equation} 
Then the twisting element $F$ defines a deformed Hopf algebra $\mathcal{A}^{(F)}:=\mathcal{A}(A;m,\Delta^{(F)},S^{(F)},\varepsilon$) with the new deformed coproduct and antipode are given as follows
\begin{equation}\label{qu23}
\begin{array}{rcl}
&&\Delta^{(F)}(X)=F\Delta(X)F^{-1},
\\[5pt]
&&S^{(F)}(X)=u^{(F)}S(X)(u^{(F)})^{-1}
\end{array}
\end{equation}
for any $X\in\mathcal{A}$, where $\Delta(X)$ and $S(X)$ are the coproduct and the antipode before twisting, and
\begin{eqnarray}\label{qu24}
&&u^{(F)}=m({\rm id}\otimes S)(F)=\sum_{i,j}f_{i}^{(1)}S(f_{j}^{(2)}),
\end{eqnarray}
if $F=\sum_{i,j}f_i^{(1)}\otimes f_j^{(2)}$. 

Let $\omega\in{A}$ be an arbitrary invertible element, $\omega\omega^{-1}=\omega^{-1}\omega=1$, then it is not difficult to check that the two-tensor
\begin{eqnarray}\label{qu25}
&&F_{\omega}:=(\omega\otimes\omega)F\Delta(\omega^{-1})
\end{eqnarray}
satisfies also the cocycle equation (\ref{qu21}). Two deformed Hopf algebras $\mathcal{A}^{(F)}=\mathcal{A}(A;m,\Delta^{(F)},S^{(F)},\varepsilon$) and $\mathcal{A}^{(F_{\omega})}=\mathcal{A}(A;m,\Delta^{(F_{\omega})}, S^{(F_{\omega})},\varepsilon$) with the twists $F$ and $F_\omega$ are isomorphic by the conjugacy isomorphism: $X\rightarrow\omega{X}\omega^{-1}$, ($\forall X\in A$). Evidently, that $\Delta^{(F)}(X)\rightarrow\Delta^{(F_{\omega})}(\omega{X}\omega^{-1})=(\omega\otimes\omega)\Delta^{(F)}(X)(\omega^{-1}\otimes\omega^{-1})$. As an example of one-parameter isomorphism family $\omega(t)$ we can use the following function:
\begin{eqnarray}\label{qu26}
&&\omega(t)=u(t):=(u^{(F)})^{t}\quad (t\in\mathbb{R}),
\end{eqnarray}
where $u^{(F)}$ is the convolution (\ref{qu24}). Therefore we have one-parameter family of the twist $F_{u(t)}$, where $F_{u(0)}=F$, $F_{u(1)}=F_{u^{(F)}}=(u^{(F)}\otimes u^{(F)})F\Delta((u^{(F)})^{-1})$ (see \cite{To2007}, \cite{BoMeMePo2019}).

Let $R$ be an universal $R$-matrix of $\mathcal{A}$ then the operator\footnote{Here and elsewhere any two tensors $F^{12}$ and $F^{21}$ are connected by the relation $F^{21}=\tau\circ F^{12}$, where $\tau$ is the superflip operator (see \ref{rmr4}).}
\begin{eqnarray}\label{qu27}
&&R^{(F)}=F^{21}RF^{-1}
\end{eqnarray}
is also the universal $R$-matrix of the twisted Hopf (super) algebra $\mathcal{A}^{(F)}$, i.e. it satisfies the relations of the form (\ref{qu5}), (\ref{qu6}). 

Let a invertible two-tensor ${\Phi}\in{A}\otimes{A}$ also generates a twist quantization after the quantization by the twist $F$, that is it satisfies the 2-cocycle condition: 
\begin{equation}\label{qu28}
\Phi^{12}(\Delta^{(F)}\otimes{\rm id})(\Phi)=\Phi^{23}({\rm id}\otimes\Delta^{(F)})(\Phi),
\end{equation}
and the normalization (\ref{qu22}), then the new co-product $\Delta^{((F)\Phi)}$ and antipode $S^{((F)\Phi)}$ are given as follows
\begin{equation}\label{qu29}
\begin{array}{rcl}
&&\Delta^{((F)\Phi)}(X)={\Phi}\Delta^{(F)}(X){\Phi}^{-1},
\\[5pt]
&&S^{((F)\Phi)}(X)=u^{((F)\Phi)}S^{(F)}(X)(u^{((F)\Phi)})^{-1}
\end{array}
\end{equation}
for any $X\in\mathcal{A}$, where $\Delta^{(F)}(X)$ and $S^{(F)}(X)$ are the coproduct and the antipode after twisting with $F$, and
\begin{eqnarray}\label{qu30}
&&u^{((F)\Phi)}=m({\rm id}\otimes S^{(F)})\Phi=\sum_{i}\phi_{i}^{(1)}S^{(F)}(\phi_{j}^{(2)}),
\end{eqnarray}
if $\Phi=\sum_{i,j}\phi_i^{(1)}\otimes\phi_j^{(2)}$. 
It is not hard to see that the invertible tensor ${\cal F}=\Phi F$ satisfies the conditions (\ref{qu21}) and (\ref{qu22}). Indeed, if we multiply  the relation (\ref{qu30}) by the relation (\ref{qu21}) with the right-hand side and apply the definition (\ref{qu23}) for $\Delta^{(F)}$ we obtain the 2-cocycle condition for ${\cal F}=\Phi F$. The normalization condition for ${\cal F}$ is obvious. Therefore the element ${\cal F}$  generates the quantization and moreover this quantization is equivalent to one in the beginning with the twist $F$ and then with $\Phi$, that is the following formulas are valid:
\begin{equation}\label{qu31}
\begin{array}{rcl}
&&\Delta^{((F)\Phi)}(X)=\Delta^{(\Phi F)}(X),
\\[5pt]
&&S^{((F)\Phi)}(X)=S^{(\Phi F)}(X).
\end{array}
\end{equation}
The first relation in (\ref{qu31}) is obvious and the second relation for the antipodes is a direct consequence of the equation for the convolutions (\ref{qu24}), (\ref{qu30}):
\begin{equation}\label{qu32}
u^{((F)\Phi)}u^{(F)}=u^{(\Phi F)},
\end{equation}
that is proved by direct calculations. It is evident that the universal $R$-matrices of these quantizations also coincide
\begin{eqnarray}\label{qu33}
&&R^{((F)\Phi)}=\Phi^{21}R^{(F)}\Phi^{-1}={\cal F}^{21}R{\cal F}^{-1}=R^{({\cal F})},
\end{eqnarray}
where ${\cal F}=\Phi F$. 

Let $\mathcal{A}^{\divideontimes}$ be a $\divideontimes$-Hopf (super)algebra, $\mathcal{A}^{\divideontimes}:=\mathcal{A}(A;m,\Delta,S,\varepsilon,\divideontimes)$, and the twisting element $F$ is unitary  
\begin{eqnarray}\label{qu34}
&&F^{\divideontimes\otimes\divideontimes}=F^{-1},
\end{eqnarray}
then the new twisting deformed $\divideontimes$-Hopf (super)algebra ${\mathcal{A}^{\divideontimes}}^{(F)}$ is also a $\divideontimes$-Hopf (super)algebra, i.e.
\begin{eqnarray}\label{qu35}
\begin{array}{rcl}
&&(\Delta^{(F)}(X))^{\divideontimes\otimes\divideontimes}=\Delta^{(F)}(X^{\divideontimes}),
\\[5pt]
&&(S^{(F)}(X))^{\divideontimes}=(S^{(F)})^{-1}(X^{\divideontimes})
\end{array}
\end{eqnarray}
for any $X\in\mathcal{A}^{\divideontimes}$. Now we come back to our concrete Hopf superalgebras $A=U(\mathfrak{osp}(1|2;\mathbb{C}))$ and ${A}^{\divideontimes}=U(\mathfrak{osp}^{\dag}(1|\mathfrak{su}(2,\mathbb{R})))$. 

{
\textit{2a. Jordanian deformation}}. There are two well-known expressions of the twisting operator $F=F_{J}$ corresponding the Jordanian classical $r$-matrix (\ref{rm30}), (\ref{is13}):
\begin{eqnarray}\label{qu36}
&&\begin{array}{rcl}
&&F_{J}=(1+1\otimes\beta E_{+})^{H\otimes1}
\\[4pt]
&&\phantom{aa}=\displaystyle1+\sum_{k>0}\frac{\beta^k}{k!}H(H-1)\cdots(H-k+1)\otimes E_{+}^k
\end{array}
\\[3pt]
&&\begin{array}{rcl}\label{qu37}
&&\phantom{aa}=\exp(H\otimes2\sigma),
\end{array}
\end{eqnarray}
where $2\sigma:=\ln(1+\beta E_{+})$ and $\beta\in\mathbb{C}$ for $U(\mathfrak{osp}(1|2;\mathbb{C}))$ and $\beta\in\imath\mathbb{R}_+$ for $U(\mathfrak{osp}^{\dag}(1|\mathfrak{su}(2,\mathbb{R})))$. The Jordanian twist in the form of binomial series (\ref{qu36}) was first proposed in \cite{GeGiSc1992} (see also \cite{KhStTo1998, KhStTo2001}) and the exponential expression (\ref{qu31}) was proposed in \cite{Og1993} (see also \cite{KuLyMu1999}). Using the binomial series (\ref{qu36}) we can easy obtain the explicit form of the element (\ref{qu24}):
\begin{eqnarray}\label{qu38}
&&u^{(F)}=m({\rm id}\otimes S)(F_J)=\exp(-\beta HE_{+}).
\end{eqnarray}
In accordance with (\ref{qu26}) we put
\begin{eqnarray}\label{qu39}
&&u(t)=\exp(-t\beta HE_{+})\quad (t\in\mathbb{R}).
\end{eqnarray}
One can calculate the following formulae for the deformed coproducts $\Delta^{(F_{J})}(X)=F_{J}\Delta(X)F_{J}^{-1}$ (see \cite{CeKu1998}):
\begin{eqnarray}\label{qu40}
\begin{array}{rcl}
&&\Delta^{(F_{J})}(e^{\pm\sigma})=e^{\pm\sigma}\otimes e^{\pm\sigma},
\\[4pt]
&&\Delta^{(F_{J})}(E_{+})=E_{+}\otimes e^{2\sigma}+1\otimes E_{+},
\\[4pt]
&&\Delta^{(F_{J})}(H)=H\otimes e^{-2\sigma}+1\otimes H,
\\[4pt]
&&\Delta^{(F_{J})}(v_{+})=v_{+}\otimes e^{\sigma}+ 1\otimes v_{+},
\\[4pt]
&&\Delta^{(F_{J})}(v_{-})=v_{-}\otimes e^{-\sigma}+1\otimes v_{-}+\beta H\otimes v_{+}e^{-2\sigma},
\end{array}
\end{eqnarray}
The coproducts $\Delta^{(F_{J})}(E_{-})$ can be calculated from the condition $\Delta^{(F_{J})}(E_{-})=4\Delta^{(F_{J})}(v_{-})\Delta^{(F_{J})}(v_{-})$. 

Using (\ref{qu23}) and (\ref{qu38}), one gets the formulas for the deformed antipode $S^{(F_J)}$:
\begin{eqnarray}\label{qu41}
\begin{array}{rcl}
&&S^{(F_{J})}(e^{\pm\sigma})=e^{\mp\sigma},\quad S^{(F_{J})}(E_{+}^{})=-E_{+}e^{-2\sigma},
\\[4pt]
&&S^{(F_{J})}(H)=-He^{2\sigma},\quad S^{(F_{J})}(v_{+})=-v_{+}e^{-\sigma},
\\[4pt]
&&S^{(F_{J})}(v_{-}^{})=-v_{-}e^{\sigma}+\beta H v_{+}e^{\sigma}.
\end{array}
\end{eqnarray}
It is easy to see the universal $R$-matrix $R^{(F)}$ for this twisted deformation looks as follows
\begin{eqnarray}\label{qu42}
R^{(F_J)}\!\!&=\!\!&F_J^{21}F_J^{-1},\quad (R^{(F_J)})^{*}=(R^{(F_J)})^{-1}.
\end{eqnarray}
In the limit $\alpha\rightarrow0$  we obtain for the $R$-matrix (\ref{qu23})
\begin{eqnarray}\label{qu43}
\begin{array}{rcl}
R^{(F_J)}\!\!&=\!\!&1+r_{J}^{}+\textit{O}(\alpha^{2}),
\end{array}
\end{eqnarray}
where $r_{J}^{}$ is the classical Jordanian $r$-matrix (\ref{is13}). It should be added that the coproduct  $\Delta^{(F_{J})}$ is real under the involution (${^\dag}$), i.e.
\begin{equation}\label{qu44}
\Delta^{(F_{J})}(a^\dag)=(\Delta^{(F_{J})}(a))^{\dag\otimes\dag},
\end{equation} 
and the antipode $S^{(F_{J})}$ satisfies the consistency
\begin{equation}\label{qu45}
S^{(F_{J})}((S^{(F_{J})}(a^{\dag}))^{\dag})=a,
\end{equation}
as well as $\epsilon(a^\star)=\overline{\epsilon(a)}$ is trivially valid for $\forall a\in U_{q}(\mathfrak{osp}^{\dag}(1|\mathfrak{sl}(2;\mathbb{R})))$. 

{
\textit{2b. Super-Jordanian deformation}}. The explicit twisting operator $F=F_{sJ}$ corresponding the super-Jordanian classical $r$-matrix (\ref{rm31}), (\ref{is14}) was obtained in the paper \cite{BoLuTo2003} and it has the following factorized form: 
\begin{equation}\label{qu46}
F_{sJ}={\Phi}F_{s}F_{J},
\end{equation}
where $F_{J}$ is the Jordanian twisting two tensor (\ref{qu37}), and the supersymmetric part $F_{s}$ depending on the odd generator $v_{+}$ and the unitarizing factor $\Phi$ are given by the formulas:
\begin{equation}\label{qu47}
F_{s}=1-4\beta\frac {v_{+}}{e^{\sigma}+1}\otimes\frac{v_{+}}{e^{\sigma}+1},
\end{equation}
\begin{equation}\label{qu48}
\Phi=\sqrt{\frac{(e^\sigma+1)\otimes(e^\sigma+1)}{2(e^\sigma\otimes e^\sigma+1)}},
\end{equation}
\begin{equation}\label{qu49}
F_{s}^{\star}=F_{s}^{-1}\,\quad {\rm for}\,\, \star=\dag\,\,{\rm or}\,\,\ddag\,,
\end{equation}
Such choice will modify the coproduct, $\Delta_{sJ}=\Phi\Delta_{sJ}\Phi^{-1}$, and we obtain
\begin{eqnarray}\label{qu50}
&&\begin{array}{r}
\Delta^{(F_{sJ})}(h)=h\otimes e^{-2\sigma}+1\otimes h+
\\[4pt]
+\beta v_{+}e^{-\sigma}\otimes v_{+}e^{-2\sigma},
\end{array}
\\[4pt]\label{qu51}
&&\begin{array}{r}
\Delta^{(F_{sJ})}(v_{+})=v_{+}\otimes1+e^{\sigma}\otimes v_{+},
\end{array}
\\[7pt]\label{qu52}
&&\begin{array}{l}
\Delta^{(F_{sJ})}(v_{-})=v_{-}^{}\otimes e^{-\sigma}+1\otimes v_{-}^{}
\\[5pt]
+\displaystyle{\frac{\beta}{4}}\biggl\{\!\Bigl(\bigl\{h,e^{\sigma}\bigr\}\otimes v_{+}^{}e^{-2\sigma}\!-
\{h,v_{+}^{}\}\otimes(e^{\sigma}\!-1)e^{-2\sigma}
\\[5pt]
+2v_{+}\otimes h-\Bigl\{h,\displaystyle{\frac{v_{+}e^{-\sigma}}{e^{\sigma}\!+1}}\Bigr\}\otimes(e^{\sigma}\!-1)e^{-\sigma}
\\[5pt]
+(e^{\sigma}\!-1)\otimes\Bigl\{h,\displaystyle{\frac{v_{+}}{e^{\sigma}\!+1}}\Bigr\}\Bigr)\displaystyle{\frac{1}{e^{\sigma}\otimes e^{\sigma}\!+1}}\biggr\}.
\end{array}
\end{eqnarray}
The formulae for the antipode $S^{(F_{sJ})}$ look as follows:
\begin{eqnarray}\label{qu53}
S^{(F_{sJ})}(h)\!\!&=\!\!&-h\,e^{2\sigma}+\frac{1}{4}(e^{2\sigma}-1),
\\\label{qu54}
S^{(F_{sJ})}(v_{+})\!\!&=&\!\!-e^{-\sigma}v_{+},
\\\label{qu55}
S^{(F_{sJ})}(v_{-})\!\!&=&\!\!-v_{-}e^{\sigma}+
\beta{h}v_{+}e^{\sigma}-\frac{\beta}{4}v_{+}e^{\sigma}.
\end{eqnarray}
It is easy to see that the formulae (\ref{qu50})--(\ref{qu52}) satisfy the reality condition
$(\Delta^{(F_{sJ})}(x))^\dag=\Delta^{(F_{sJ})}(x^\dag)$ for any $x\in\mathfrak{osp}^{\dag}(1|\mathfrak{sl}(2;\mathbb{R}))$ and the antipodes (\ref{qu53})--(\ref{qu55}) satisfy the condition (\ref{qu35}). 

The universal $R$-matrix of the super-Jordanian deformation has the form
\begin{equation}\label{qu56}
R^{(F_{sJ})}={\Phi}F^{21}_{s}R^{(F_{J})}F^{-1}_{s}{\Phi}^{-1},
\end{equation}
where
\begin{equation}\label{qu57}
R^{(F_{J})}=F^{21}_{\!J}F^{-1}_{\!J}=e^{2\sigma\otimes h}\otimes e^{-2h\otimes\sigma}.
\end{equation}

\section{Short Summary and Outlook}
By the simple algebraic technique we obtain the complete classification of all basic (nonisomorphic) quantum deformations for the orthosymplectic Lie superalgebra $\mathfrak{osp}(1|2;\mathbb{C})$ and its pseudoreal and real forms in terms of the classical $r$-matrices. In particular, we prove that compact pseudoreal form has only one quantum deformation (standard $q$-analog), and the $D=3$, $N=1$ Lorentz supersymmetry, which is the non-compact real form of $\mathfrak{osp}(1|2;\mathbb{C})$, has four different Hopf-algebraic quantum deformations: two standard $q$-analogs, and two (Jordanian and super-Jordanian) twist deformations. All basic Hopf-algebraic quantum deformations are presented in the explicit form.

In conclusion of this article, we want to note a very important point related to the fact that the results obtained here will help solve a more difficult problem: classification and construction of quantum deformations for the orthosymplectic superalgebra $\mathfrak{osp}(1|4;\mathbb{C}))$ and its real (de Sitter and anti-de Sitter) forms, which are directly used in SUGRA models. Namely, the superalgebra $\mathfrak{osp}(1|4;\mathbb{C}))$, as a linear space, has the following tensor structure
\begin{equation}\label{qu58}
\mathfrak{osp}(1|4;\mathbb{C}))=\mathfrak{osp}_{L}(1|2;\mathbb{C}))\oplus{\cal{P}}_{4}\oplus\mathfrak{osp}_{R}(1|2;\mathbb{C})),  
\end{equation}
where left $\mathfrak{osp}_{L}(1|2;\mathbb{C}))$ and right $\mathfrak{osp}_{R}(1|2;\mathbb{C}))$ superalgebras are isomorphic to the one considered here $\mathfrak{osp}(1|2;\mathbb{C}))$, and ${\cal{P}}_{4}$ is a four-dimensional linear space of a curved four-momentum generators. Since we already know the quantum deformations for the left and right superalgebras, our task is to extend them to the curved four-momentum ${\cal{P}}_{4}$.

Finally it should be mentioned that one can introduce also quaternionic superalgebra $\mathfrak{osp}(1|2;\mathbb{H})$ \cite{LuNo1982,PiNiSo1985} which describes $D=5$ Lorentz or $D=4$ de-Sitter superalgebra with the bosonic sector $\mathfrak{o}(4,1)\otimes\mathfrak{o}(2)$, and consider their quantum deformations. 

\subsection*{Acknowledgments}
The author would like to thank A. Borowiec and J. Lukierski for useful discussions. The paper was supported by RFRB grant No 19-01-00726A.


\end{document}